\documentclass[lettersize,journal]{IEEEtran}
\usepackage{amsmath,amsfonts,amssymb}
\usepackage{algorithmic}
\usepackage{algorithm}
\usepackage{array}
\usepackage[caption=false,font=normalsize,labelfont=sf,textfont=sf]{subfig}
\usepackage{textcomp}
\usepackage{stfloats}
\usepackage{url}
\usepackage{verbatim}
\usepackage{graphicx}
\usepackage{cite}
\usepackage{amsmath,bm,flushend}

\makeatletter

\newcommand{\Rmnum}[1]{\expandafter\@slowromancap\romannumeral #1@}
\makeatother
\begin{document}

\title{Reconfigurable Intelligent Computational Surfaces for MEC-Assisted Autonomous Driving Networks: Design Optimization and Analysis}

\author{Xueyao Zhang, Bo~Yang, Zhiwen Yu,~\IEEEmembership{Senior Member,~IEEE}, Xuelin~Cao,  George C. Alexandropoulos,~\IEEEmembership{Senior Member,~IEEE}, Yan Zhang,~\IEEEmembership{Fellow,~IEEE},  %Yuanwei Liu,~\IEEEmembership{Fellow,~IEEE},
 M\'erouane Debbah,~\IEEEmembership{Fellow,~IEEE},  %H. Vincent Poor,~\IEEEmembership{Life Fellow,~IEEE}, 
 and Chau Yuen,~\IEEEmembership{Fellow,~IEEE}%, Lajos Hanzo,~\IEEEmembership{Life Fellow,~IEEE}, and
        % <-this % stops a space
\thanks{X. Zhang, B. Yang, and Z. Yu are with the School of Computer Science, Northwestern Polytechnical University, Xi'an, Shaanxi, 710129, China. X. Cao is with the School of Cyber Engineering, Xidian University, Xi'an, Shaanxi, 710071, China. G. C. Alexandropoulos is with the Department of Informatics and Telecommunications, National and Kapodistrian University of Athens, 15784 Athens, Greece. Y. Zhang is with the Department of Informatics, University of Oslo, 0316 Oslo, Norway.  
M. Debbah is with the Center for 6G Technology, Khalifa University of Science and Technology, P O Box 127788, Abu Dhabi, United Arab Emirates. 
C. Yuen is with the School of Electrical and Electronics Engineering, Nanyang Technological University, Singapore. 
 
  (\textit{Corresponding author:} Bo Yang).}% <-this % stops a space
}

\markboth{Journal of \LaTeX\ Class Files,~Vol.~xx, No.~xx, 2024}%
{Shell \MakeLowercase{\textit{et al.}}: A Sample Article Using IEEEtran.cls for IEEE Journals}


\maketitle

\begin{abstract}
This paper investigates autonomous driving safety improvement via task offloading from cellular vehicles (CVs) to a multi-access edge computing (MEC) server using vehicle-to-infrastructure (V2I) links. Considering that the latter links can be reused by vehicle-to-vehicle (V2V) communications to improve spectrum utilization, the receiver of the V2I link may suffer from severe interference that can cause outages during the task offloading. To tackle this issue, we propose the deployment of a reconfigurable intelligent computational surface (RICS) whose computationally capable metamaterials are leveraged to jointly enable V2I reflective links as well as to implement interference cancellation at the V2V links. We devise a joint optimization formulation for the task offloading ratio between the CVs and the MEC server, the spectrum sharing strategy between V2V and V2I communications, as well as the RICS reflection and refraction matrices to maximize an autonomous driving safety task. Due to the non-convexity of the problem and the coupling among its free variables, we transform it into a more tractable equivalent form, which is then decomposed into three sub-problems solved via an alternate approximation method. Our simulation results showcase that the proposed RICS-assisted offloading framework significantly improves the safety of the considered autonomous driving network, yielding a nearly 34\% improvement in the safety coefficient of the CVs. In addition, it is demonstrated that the V2V data rate can be improved by around 60\% indicating that the RICS-induced adjustment of the signals can effectively mitigate interference at the V2V link.
\end{abstract}

\begin{IEEEkeywords}
Reconfigurable intelligent computational surfaces, autonomous driving,  multi-access edge computing, spectrum sharing, task offloading.
\end{IEEEkeywords}

\section{Introduction}
\begin{figure}[t]
	\centering
	\includegraphics[width=0.9\columnwidth]{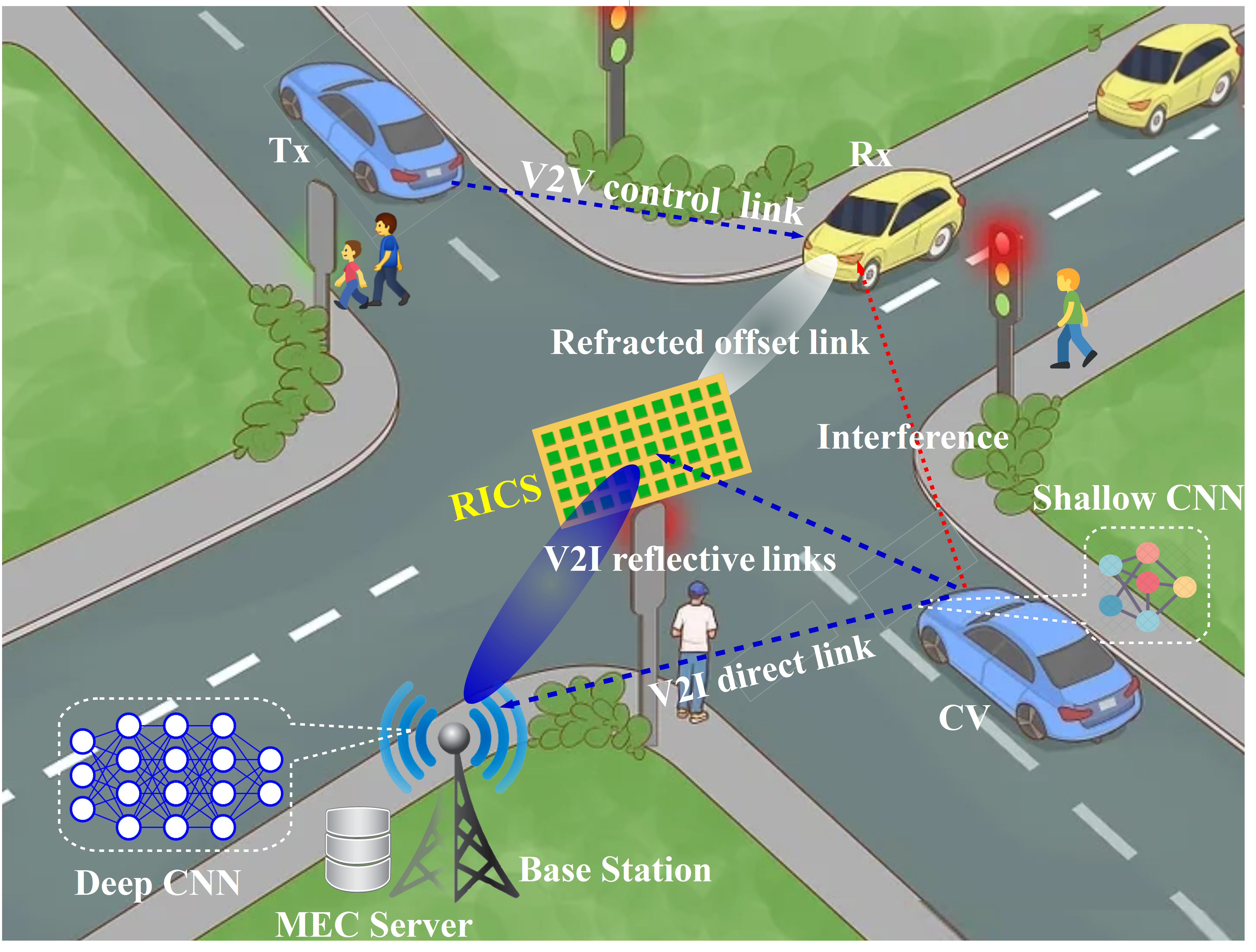}
	\caption{The considered RICS-aided autonomous driving paradigm, where a V2V Rx experiences severe co-channel interference from neighboring CVs. An adequately optimized RICS can mitigate the interference of the V2V link, while improving the V2I link performance.}
    \label{system}
\end{figure}

\IEEEPARstart{A}{long} with the exponential evolution of wireless technologies, next-generation mobile networks will deliver low-latency and high-reliability connectivity for intelligent vehicular transportation systems \cite{B5G}. The booming adoption of artificial intelligence (AI) in industrial automation applications is driving extensive consideration of deep learning (DL) techniques to improve safety in autonomous driving scenarios. As on-board sensors can generate huge amounts of multi-modal data, their efficient exploitation for decision-making within a limited time becomes a challenge. In this context, edge intelligence (EI) has become critical to enable the processing of data uploaded by autonomous vehicles, and then make decisions at the nearby multi-access edge computing (MEC) server \cite{YB_wc,EI01,EI02}. To further boost the safety of autonomous driving, vehicle-to-infrastructure (V2I) and vehicle-to-vehicle (V2V) wireless links for realizing computation offloading and sharing critical safety information between vehicles are becoming significant. However, as shown in Fig.~\ref{system}, the V2V receiver `Rx' may suffer from severe co-channel interference caused by neighboring cellular vehicles (CVs), which contributes to autonomous driving safety. 

\vspace{-2mm}
\subsection{Motivation and Scope}
 \subsubsection{Accidents with Autonomous Driving} According to the National Transportation Safety Board (NTSB) accident report, a Tesla Model S vehicle equipped with the "Autopilot suite" struck a refrigerated semi-trailer driven by a tractor-trailer on US Highway 27A (US-27A) in Florida at 4:36 pm Eastern daylight time on 7 May 2016~\cite{NTSB1}. But less than two years later, an Uber self-driving test vehicle based on a modified 2017 Volvo XC90 collided with a pedestrian who was crossing the road with a bicycle in Arizona at around 9:58 p.m. on March 18, 2018~\cite{NTSB}.

Abide by the basic ecosystem of an autonomous vehicle, the self-driving vehicles involved in the accidents were mainly equipped with forward- and side-facing cameras, radars, light detection and ranging (LIDAR), navigation sensors, and a computing and data storage unit that performs online inference. In particular, according to data from the self-driving system in the Uber test vehicle, the pre-trained DL model mistakenly classified the pedestrian as an unknown object, then as a vehicle, and finally as a bicycle with varying expectations of the future trajectory. Due to the constant erroneous inference results given by the pre-trained DL model, the time cost to achieve the correct inference is increased. In the accident, only about $1.3$ seconds were left for the vehicle to brake before impact, which is far from enough\footnote{In the Uber accident, the vehicle was traveling at about $43$ mph before braking. If the vehicle was in a lower speed range or the vehicle was slowed down beforehand, the accident could be avoided}. In this context, {how to achieve ``fast and accurate inference" for improving autonomous driving safety is the goal of this paper}.

%To improve the wireless propagation link quality, reconfigurable intelligent surfaces (RISs) become a promising technology \cite{RIS00}, thus, they can also assist vehicular communications \cite{RIS01}, \cite{RIS02}. However, considering a typical scenario illustrated in Fig.~\ref{scene}(b), V2V receivers (e.g., Rx) may suffer from severe co-channel interference caused by neighboring cellular vehicles (CVs), which contributes against driving safety. To combat interference at the Rx side of the V2V control link, conventional RISs fall short short in providing significant interference suppression needed for autonomous driving \cite{RIS03}. To solve this issue, we consider, in this paper, reconfigurable computational intelligent surfaces (RICSs) \cite{RICS} that are capable of both phase-shifting and amplifying their impinging signals directly in the analog domain, and present a novel framework for their efficient optimization for computation offloading of CVs, while suppressing interference in V2V communication pairs.

Reconfigurable intelligent surfaces (RISs) play a significant role in improving the quality of wireless links~\cite{RIS00,RIS_amp}, and can therefore support vehicular communications as well~\cite{RIS01, RIS02}. However, to combat interference on the receiver side of the V2V control link, conventional RISs fail to provide significant interference suppression that is needed for autonomous driving~\cite{RIS03}. This happens due to their zero or minimal computing capabilities. To address this problem, in this paper we introduce Reconfigurable Intelligent Computing Surfaces (RICSs)~\cite{RICS}, which are capable of both phase shifting and adjusting their impinging signals directly in the analog domain, and then we present a novel framework for their efficient optimization for computation offloading of CVs while suppressing interference in V2V communication pairs. In contrast to the RIS-aided MEC approaches in~\cite{RISmec01,RISmec02,RISmec03}, the proposed scheme optimizes the RICS structure to perform specific computational tasks, rather than solely using the traditional RIS paradigm to improve the offloading link budget via optimized reflective beamforming. 

Motivated by the above, this paper focuses on how to improve the safety of autonomous driving through vehicle computation offloading assisted by RICS. By offloading the computation tasks of vehicles to the MEC server through V2I links, the computational capabilities of vehicles can be extended and thus the risks of autonomous driving can be reduced. Meanwhile, by appropriately configuring the RICS, the mitigation of interference in V2V communication can be achieved as well.

\subsection{Related Works}
Recent studies have highlighted the importance of introducing advanced wireless communication technologies (such as B5G) to improve the safety of autonomous driving. Specifically, the authors of \cite{autosafety01} explore the application of wireless transmission technology to improve the safety and reliability of autonomous driving systems, with a focus on secure communication and remote monitoring. The authors of \cite{autosafety02} investigate the challenges and solutions in network communication for autonomous driving systems, and propose an adaptive fusion engine to mitigate the impact of network latency fluctuations on the reliability and safety of autonomous vehicles. In addition, the authors of \cite{V2X01} demonstrate that the increased bandwidth in B5G networks facilitates faster and more reliable V2X communication, which is critical for real-time safety applications.  In \cite{V2X02} the authors explore the key role of 5G in connected and cooperative autonomous driving and discuss the use of machine learning to enhance V2X services. The authors of ~\cite{V2X03} present a collaborative autonomous driving framework based on C-V2X, which provides safety for the communications between vehicles. 

Meanwhile, most of the existing literature on offloading in-vehicle networks aims at uploading computationally intensive tasks to the edge servers to improve computational efficiency~\cite{MEC_vehicle}. It is well known that MEC has become an effective solution for processing sensing data from autonomous vehicles, thereby reducing dependence on central cloud servers and enabling real-time data analysis~\cite{EdgeComp01}. In particular, the authors of \cite{EI01} explore the complexity of autonomous driving systems and highlight the challenges faced by edge computing systems. It also emphasizes the importance of vehicle-to-everything (V2X) technology in providing redundancy and alleviating the stringent requirements of edge computing workloads. In \cite{EdgeComp03}, the authors investigate the relationship between edge computing and vehicle safety and design an intelligent vehicle-road cooperative system using mobile edge computing to decrease the network latency of data transmission. In addition, \cite{EdgeComp04} proposes a MEC scheme that migrates computing and storage capabilities to network edge nodes to meet the requirements of performing computationally intensive or delay-sensitive tasks on vehicles. %It proposes a cooperative offloading decision-making method using offloading games to ensure the security of computation offloading between vehicles.

Notably, DL techniques at the edge can help to deeply explore the inherent characteristics of the collected huge dataset from heterogeneous sensors and make more reasonable decisions in near-real-time. Among the related contributions, \cite{DL01} proposes to predict the road traffic situation using the convolutional neural network (CNN), and then a proactive load balancing approach is presented to enable cooperation among mobile edge servers. The authors of~\cite{DL02} investigate the characteristics of a Rayleigh fading channel and propose to train a long short-term memory (LSTM) model to predict future channel parameters. In~\cite{DL03} the authors implement the classical supervised machine learning methods for detecting the non-line-of-sight (NLoS) conditions by learning the V2V measurement data. However, none of the above works consider the impact of the collected data quality on the inference accuracy of the trained DL model, which is the critical trigger for autonomous driving accidents, such as Uber and Tesla.

Finally, in order to improve wireless transmission performance, there exists a large number of researchers investing in RIS-assisted wireless communication. In \cite{RIS-ass01}, the authors investigate the communication technology in RIS-assisted MEC system and propose a reinforcement learning-based power allocation optimization algorithm to improve the safety of data transmission.  In \cite{RIS-ass02}, the authors propose a block coordinate descent algorithm to minimize offloading latency in RIS-enabled vehicular networks. A joint optimization of RIS reflection coefficients, spectrum sharing~\cite{SS1, SS2}, and MUD matrix is investigated in \cite{QoS}. Numerous studies have been carried out on STAR-RIS with simultaneous transmission and reflection capabilities. In \cite{STAR-ass01}, a STAR-RIS assisted V2X communication system is studied where joint optimization is employed to maximize the achievable data rate for V2I users while satisfying the latency and reliability requirements of V2V pairs. In \cite{STAR-ass02}, the authors discuss the communication performance of V2V networks using RIS and STAR-RIS under non-orthogonal multiple access (NOMA) and orthogonal multiple access (OMA) schemes. Experiments show that the application of RIS/STAR-RIS technology in V2V communication significantly improves the communication performance of intelligent transportation systems (ITS). In addition, multi-layer RISs called Stacked Intelligent Metasurfaces (SIM) can perform signal processing directly in the electromagnetic wave domain by stacking multiple metasurface layers, which provides better performance than single-layer metasurfaces. In \cite{sims}, the authors proposed a SIM-based holographic multiple-input multiple-output (HMIMO) communication system to realize beamforming and demodulation. Compared to the aforementioned literature, our proposed RICS-based offloading scheme is able to carry out specific computational tasks, rather than solely using the traditional RIS paradigm to improve the offload link budget via optimized reflective beamforming. 

\subsection{Contributions and Organization}
The contributions of this paper are summarized as follows.

\begin{itemize} 
\item We introduce a novel RICS structure by enabling computation functions via metamaterials. In particular, by fully exploiting the computational capability of RICS, the intelligence computational layer can be configured to adjust the impinging signal's amplitude dynamically, and thus the interference suffered at the receiver of the V2V pair can be mitigated. Based on this, we further propose an RICS-aided MEC framework for autonomous driving. Since the inference delay and accuracy are interplaying due to the diversity of the model capabilities implemented at the vehicles and MEC server, a joint optimization problem on offloading ratio, spectrum sharing strategy and RICS's reflection and refraction matrices is formulated to maximize the safety coefficient of the CVs while satisfying the outage probability of V2V pairs, thereby enhancing the autonomous driving safety. 

\item Consider that the variables to be optimized are interrelated with each other, the formulated optimization problem is mixed integer nonlinear programming (MINLP) and thus is difficult to obtain the optimal solutions. To solve this problem efficiently, we decompose the original problem into three sub-problems, which are jointly solved through the proposed alternating iterative optimization algorithm (AIOA) alternately until convergence is achieved. Meanwhile, in order to find the optimal amplitude adjustment factors of the analog computing layer to effectively mitigate interference on V2V Rx, a quadratic optimization (QP) problem was constructed by transforming the original optimization problem through least squares, the optimal amplitude adjustment factors are obtained using the gradient descent method.

\item Finally, we conduct the performance evaluation by comparing several benchmarks to verify the superiority of the proposed RICS-aided MEC framework. With the optimized offloading strategy scheme, our proposed framework outperforms the other benchmark schemes in terms of the safety coefficient of V2I links and the sum data rate of V2V links.
\end{itemize}

\textit{Notations}:  
The following are mathematical operation symbols involved in this paper: ${\bm{A}^{H}}$ stands for performing the conjugate transpose operation on the matrix $\bm{A}$. $|\bm{A}|$ represents the absolute value. The real and imaginary parts of complex numbers are denoted by $\Re(\cdot )$ and $\Im(\cdot )$ respectively. $Tr(\bm{A})$ stands for the trace of the matrix $\bm{A}$. $rank(\bm{A})$ is the rank of the matrix $\bm{A}$. $diag(\cdot )$ denotes the diagonalization of a vector. The argument of a complex number is denoted by $\angle$. $\bm{A}\circ \bm{B}$ is the element-wise multiplication of two matrices $\bm{A}$ and $\bm{B}$ of the same dimension. $j$ is an imaginary unit satisfying the equation ${{j}^{2}}=-1$ and ${{\mathbb{C}}^{m\times n}}$ is a complex matrix.

The paper is organized as follows. In Section~\ref{sention_2} we present the structure of RICS and illustrate the principle for achieving amplitude adjustment via metamaterials. Section~\ref{sention_3} presents the RICS-assisted autonomous driving scenario and the system models, followed by the problem formulation and analysis. Section~\ref{sention_5} introduces an alternative iterative optimization algorithm to solve the formulated problem, where the optimal amplitude adjustment factors are explored and the complexity analysis is given as well. In Section~\ref{sention_6}, we evaluate and discuss the simulation results to demonstrate the advantages of the proposed algorithm. Finally, the whole paper is summarized in Section~\ref{sention_7}.

%\begin{figure*}[!t]
%\centering
%\subfloat[The structure of the proposed RICS working in RR+AC mode]{\includegraphics[width=2.5in]{Fig1b.pdf}%
%\hfil
%\subfloat[RICS-aided internet of autonomous vehicular networks]{\includegraphics[width=2.5in]{Fig1.png}%
%\label{fig_second_case}}
%\caption{The proposed RICS-aided autonomous driving paradigm is shown in (a), where a V2V Rx suffers severe co-channel interference from neighboring CVs. An optimized RICS can mitigate the interference of the V2V link, while improving the V2I link performance. In (b), the RICS structure is configured as RR+AC mode, being capable to create an ``interference-free
%zone" via properly configuring the relative permittivity and permeability of the metamaterial included in the intelligence computation layer.}
%\label{scene}
%\end{figure*}

\section{RICS Design Fundamentals}\label{sention_2}
%\subsubsection{Mobile edge computing for vehicular networks}
%\subsubsection{Reconfigurable intelligent surfaces aided vehicular communications}
%Existing literature primarily focuses on offloading computation-intensive tasks to edge servers in vehicle networks to enhance computing efficiency~\cite{MEC_vehicle}. Besides, DL techniques at the edge may help deeply dig up the inherent characteristics of the collected big data from heterogeneous sensors and make more reasonable decisions in the vehicle networks.
%In this part, we explore the crucial research advances in enhancing the safety and efficiency of autonomous driving.

\subsection{RICS Structure} 
\begin{figure}[t]
	\centering
	\includegraphics[width=0.95\columnwidth]{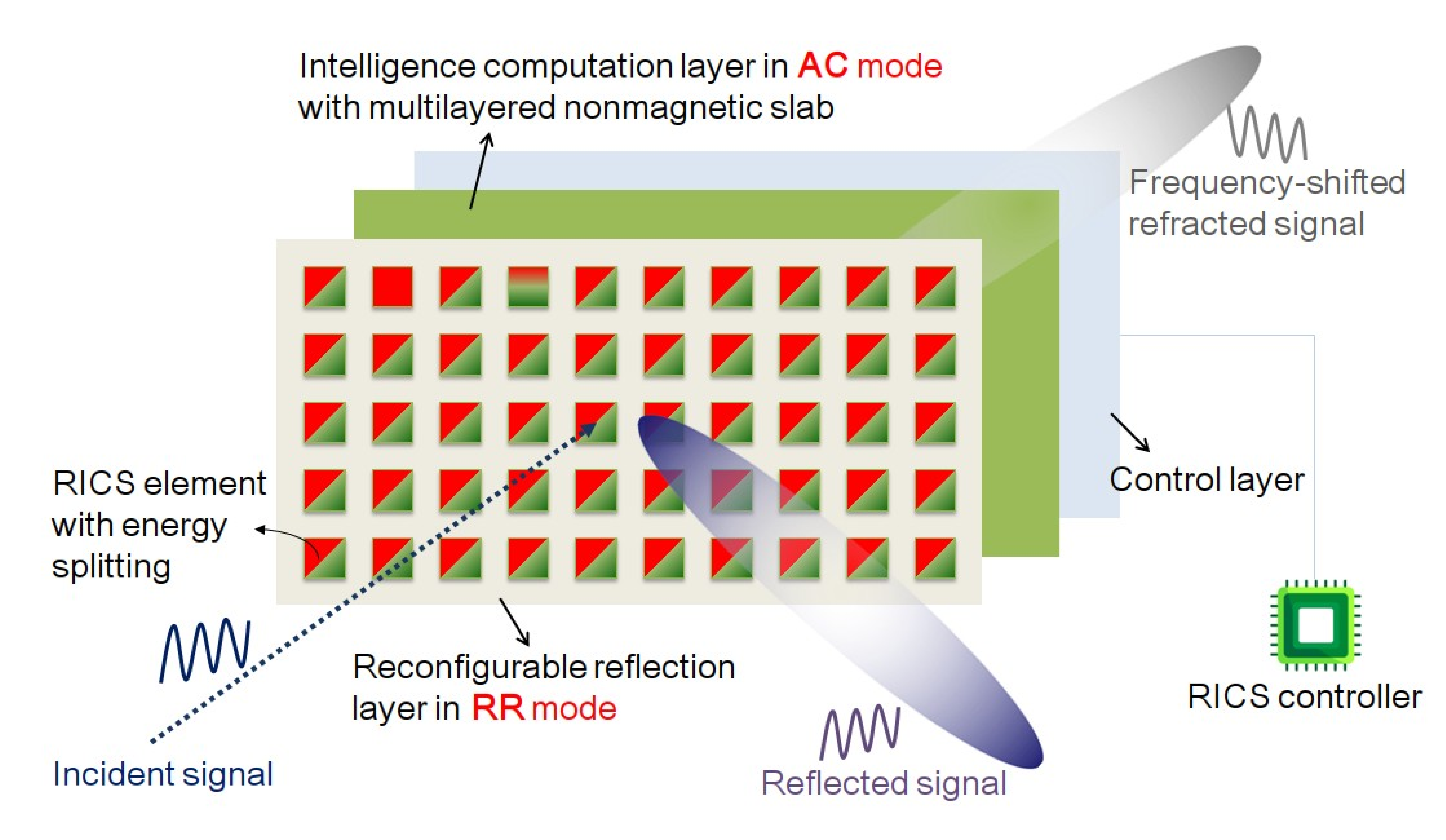}
	\caption{The structure of the proposed RICS working in RR+AC mode, being capable to create an ``interference-free
zone" via properly configuring the relative permittivity and permeability of the metamaterial included in the intelligence computation layer.}
    \label{RICS}
\end{figure}

As conceptually sketched in Fig.~\ref{RICS}, an RICS consists of a control layer and two functional layers: the \textbf{reconfigurable beamforming layer} and the \textbf{intelligence computation layer}, which can be jointly configured by an intelligent controller. Here, the reconfigurable reflection layer is configured as the \textbf{reflection-refraction (RR) mode} and the intelligence computation layer is configured as the \textbf{analog-computing (AC) mode}~\cite{RICS}. To improve signal coverage, the signal incident on each element can be split into two parts: some of the energy is used for signal reflection, while the remaining energy can support signal refraction to serve users located on the opposite side~\cite{omni, CoupledPhase}. 

Different from these existing RIS designs, e.g., STAR-RIS, our proposed RICS introduces the two layer structure, which gives it more powerful functions and flexibility. The reconfigurable beamforming layer can work not only in reflection-absorption (RA) mode, which realizes the dynamic adjustment of the incident RF signal by reflecting part of the signal and absorbing the other part of the signal, but also in RR mode, where the incident signal is divided into reflected and refracted signals, and the refracted signals are adjusted by the intelligence computation layer with analog-computing mode. Focusing on the application scenario of this paper, i.e. autonomous driving safety, it is desirable to counteract the interference via the refracted phase-shifted signal. However, since the refracted signal is very limited,  in this paper we consider utilizing the intelligence computation layer of the RICS to perform \textit{amplitude adjustment} on the phase-shifted refracted signals by configuring the tunable metamaterial parameters, aiming at interference suppression at the receiver of the V2V pair. With its innovative double-layer structure and diverse operating modes, our RICS offers the flexibility to choose the right operating mode for the actual application scenario and to configure it precisely to better meet the actual needs.%The two functional layers interact and should be jointly configured.

Specifically, let $L$ denote the total number of elements of RICS, the energy splitting ratio of the $l$th element is defined as ${\chi_l} \!\triangleq\! \beta_l^{r}\!:\! \beta_l^{t}, \ \forall l \in {\cal L} \triangleq \{1, 2, \ldots, L\}$, where $\beta_l^{r}\in [0,1]$ and $\beta_l^{t} \in [0,1]$ indicate the reflection and refraction (also transmission) amplitude coefficients of each element, respectively, and $\beta_l^{r}+\beta_l^{t}\!=\!1$ generally hold. To characterize the feature of RICS, let $s_l$ denote the signal incident on the $l$th element of the RICS, then the signals reflected and refracted by the $l$th element can be presented as $s_l^{r}\!=\!\sqrt{\beta_l^{r}}e^{j \theta_l^r}s_l$  and $s_l^{t} \!=\! \Psi_{l}\sqrt{\beta_l^{t}}e^{j \theta_l^t}s_l$, where $\theta_l^r\in(0,2\pi]$ and $\theta_l^t\in(0,2\pi]$ indicate the phase shifts for reflection and refraction of each element, respectively. $\Psi_l$ represents the amplitude adjustment factor of the $l$th RICS element, and when we are configuring the RR mode, $\bm{\Psi}=1$ always hold. In addition, based on the physical characteristics of the RR mode, there exists a fixed phase difference between reflection and refraction. Therefore, we have the following relationship: $\left| {{\theta }_{t}}-{{\theta }_{r}} \right|=\frac{\pi }{2}or\frac{3\pi }{2}, 1\le l\le L$. As for the deployed RICS, the reflection and refraction coefficient matrices are given by ${\mathbf{\Phi}}_r\!=\!{\rm{diag}} \left (s_1^{r}, s_2^{r}, \ldots, s_L^{r} \right )$ and ${\mathbf{\Phi}}_t\!=\!{\rm{diag}} \left (s_1^{t}, s_2^{t}, \ldots, s_L^{t} \right )$, respectively.

 %However, this leads to the fact that there is a limit to the amount of signal energy that can be used to counteract interference as transmitted. So we considered utilizing the intelligence computation layer with AC mode to be able to perform amplification operations on the refracted signals to match the transfer function under consideration, thus eliminating the interference at the receiver of the V2V pair. 
\subsection{Amplitude Adjustment via RICS's Metamaterials} \label{sec:algorithm-three-b}
In this subsection, we have developed a metamaterials system that achieves amplitude adjustment of the incident wave by creating an enhanced signal distribution. The core of this design lies in the precise control of both the amplitude and phase of the wave, enabling constructive interference with the V2I interfering wave. Importantly, this process does not violate energy conservation, as the constructive interference effect stems not from an increase in energy, but rather from the optimized adjustment of the wave amplitude distribution, which effectively counteracts the interference from the V2I link in the spatial domain. With the help of modulation technology, metamaterials can induce the reconstruction of incident waves by changing their own structure and parameters, which realizes the concentration or dispersion of energy in the domain of metamaterials and thus achieves the effect of amplifying or weakening the signals without additional power consumption.

In the considered system, we let $\bm{\Psi} \!=\!\{{{\Psi }_{1}},\ldots,{{\Psi }_{l}}\}$ indicate the vector of amplitude adjustment factors. As the incident signal impinging on the RICS, the adjusted signal of the $l$th element can be described as 
$f(x_l) \!=\! {\Psi}_l x_l+n_s$, where ${\Psi}_l x_l$ is the desired signal and $n_s$ indicates the static noise~\cite{active}. When the value of $\Psi$ varies within $\left(0, 1\right)$, the signal is \textit{weakened}, and when it is larger than $1$, the signal will be \textit{amplified}. Considering that the static noise has nothing to do with $\Psi$, we ignore it here and thus the adjustment of the desired signal can be considered essentially as a mathematical operation. Since mathematical operations can be achieved with metamaterial design~\cite{metamaterials}, we can perform incident signal adjustment by designing the intelligence computation layer. The design architecture of this metamaterials contains three modules: the Fourier transform, the transfer function and the inverse Fourier transform. The specific roles of the three modules are as follows:
\begin{itemize}
\item
\textbf{The Fourier transform}: realizes the conversion of input signals from the spatial domain to the frequency domain and provides the basis for signal processing in the frequency domain.
\item
\textbf{The transfer function}: modulates the amplitude and phase of the signal, thereby realizing mathematical operations.
\item
\textbf{The inverse Fourier transform}: converts the processed signal from the frequency domain back to the spatial domain.
\end{itemize}

\begin{figure}[t]
	\centering	\includegraphics[width=0.6\columnwidth]{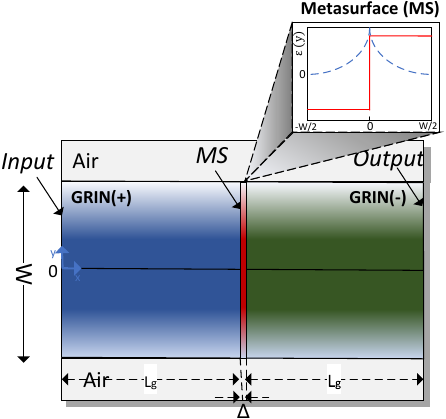}
	\caption{The combined `GRIN-MS-GRIN' structure for achieving signal amplitude adjustment.}
    \label{MS}
\end{figure}

To realize the three conceptual modules mentioned above, firstly, we use a graded refractive index (GRIN) metamaterial to achieve the Fourier transform and the inverse Fourier transform~\cite{GRIN}. Specifically, the design of the transfer function can be achieved via the metasurface (MS), e.g., consisting of Aluminum-doped ZnO (AZO) and Silicon (Si). Therefore, to realize the analog computing function for the amplitude adjustment, we implement a combined `GRIN-MS-GRIN' structure, as shown in Fig.~\ref{MS}. %In this system, a Fourier transform is performed after each propagation over a characteristic length ${{L}_{g}}$. 
By inserting an MS between two GRIN lenses, the wavefront amplitude can be modulated, thereby enabling mathematical operations on the input function. In the following, we introduce the details for achieving the three modules.  %Once the transfer function has been determined, we need to design the metamaterial parameters of the GRIN and MS.

\subsubsection{Design for achieving Fourier and inverse Fourier transforms} Let $y$ denote the coordinates of the spatial position of the electromagnetic wave during propagation, then the permittivity of the GRIN material at the coordinate $y$ can be expressed as
$\varepsilon \left( y \right)\!=\!{{\varepsilon }_{c}}\left\{ 1-{{\left[ \pi /\left( 2{{L}_{g}} \right) \right]}^{2}}{{y}^{2}} \right\}$, where the constant ${\varepsilon}_c$ denotes the permittivity at the central plane of the 2D GRIN material, $L_{g}$ denotes the characteristic distance of the GRIN block. 

Specifically, as shown in Fig.~\ref{MS}, there are two types of GRIN materials. GRIN(+) denotes a positive relative permittivity with ${\varepsilon}_{c}\!=\!2.01{\varepsilon }_{o}$ and $\mu\!=\!\mu_0$. This corresponds to the Fourier transform domain in the conceptual module described above. By contrast, GRIN(-) has a negative relative permittivity (${\varepsilon }_{c}\!=\!-2.01{\varepsilon }_{o}$ and $\mu\!=\!-\mu_0$), which corresponds to the inverse Fourier transform domain. Here, ${{\varepsilon }_{0}}=8.85\times {{10}^{-12}}F/m$ and ${{\mu }_{0}}=4\pi \times {{10}^{-7}}H/m$ respectively represent the permittivity of free space and the permeability of free space. Moreover, some parameters related to the design of GRIN materials are given as follows: ${{\lambda }_{o}}\!=\!3 \mu m$ indicates the wavelength of electromagnetic waves in free space, the characteristic length is given by ${{L}_{g}}\!=\!35 \mu m$, $W\!=\!10{{\lambda }_{o}}$ represents the width of the graded refractive index material in the transverse direction.

\subsubsection{Design for realizing transfer functions} 
The concrete realization of the mathematical operations heavily relies on the design of the transfer function, which can be realized via the MS. We denote a certain operation realized by the transfer function as $G(y)$ and the input signal at the $l$th element is $x_l(y)$, the convolution of the input function with the transfer function can be expressed as $g\left( y \right)\!=\!\int{x_l\left( u \right)G\left( y-u \right)du}$. This operation in the Fourier space can be represented as $\tilde{g}(y)\!=\!\tilde{G}(y)\tilde{x}_l(y)$, where the tilde represents Fourier transform. This indicates that when the signal ${x}_l(y)$ has traveled $L_g$ along the GRIN, we can obtain $\tilde{x}_l(y)$.

In this context, our goal becomes obtaining the transfer function $G(y)$. 
%That is to say, after the input function $f\left( y \right)$ enters the implemented transfer function module, the mathematical operations on the input function are completed. Subsequently, the frequency-domain signal resulting from these mathematical operations is passed into the inverse Fourier transform domain for conversion into a spatial domain signal. In other words, we need to find a suitable transfer function $G\left( y \right)$ based on some properties of the Fourier transform and the desired mathematical operations we want to implement. 
Since we consider adjusting the amplitude of the signal without changing its time scale, the signal entering the $l$th element of the RICS analog computation mode can be expressed as follows: 
\begin{equation} \label{eq1}
f\left(x_l(y)\right)={{\mathcal{F}}^{-1}}\left\{\left( \frac{1}{{\Psi}_l}\right) \mathcal{F}\left( x_l(y) \right)\right\}.
\end{equation}
%Different RICS elements may have different amplification factors. Obviously, according to the following equation, we can obtain the transfer function: $\overset{\sim}{\mathop{G}}\,\left( y \right)\propto \frac{1}{{{\Psi }_{l}}}$.
%After completing the selection of materials corresponding to each of the above conceptual modules,
%To realize the analog computing function in (\ref{eq1}), we implement a combined `GRIN-MS-GRIN' structure, as shown in Fig.~\ref{MS}. %In this system, a Fourier transform is performed after each propagation over a characteristic length ${{L}_{g}}$. 
%By inserting a metasurface (MS) between two GRIN lenses, the wavefront amplitude can be modulated, enabling mathematical operations on the input function. Once the transfer function has been determined, we need to design the metamaterial parameters of the GRIN and MS. 

As indicated in Fig.~\ref{MS}, the MS needs to modify the amplitude distribution that propagates a distance of $L_{g}$ along the GRIN(+). At this point, it is proportional to the first Fourier transform of the input function, i.e., $G(y)  \propto \left( \frac{1}{{{\Psi }_{l}}} \right)\mathcal{F}(x_l(y))$. %After propagating through the MS field, the output domain should be proportional to $\left( \frac{1}{{{\Psi }_{l}}} \right)\mathcal{F}(x_l(y))$.

\begin{figure}[t] 
	\centering
    \includegraphics[width=1.0\columnwidth]{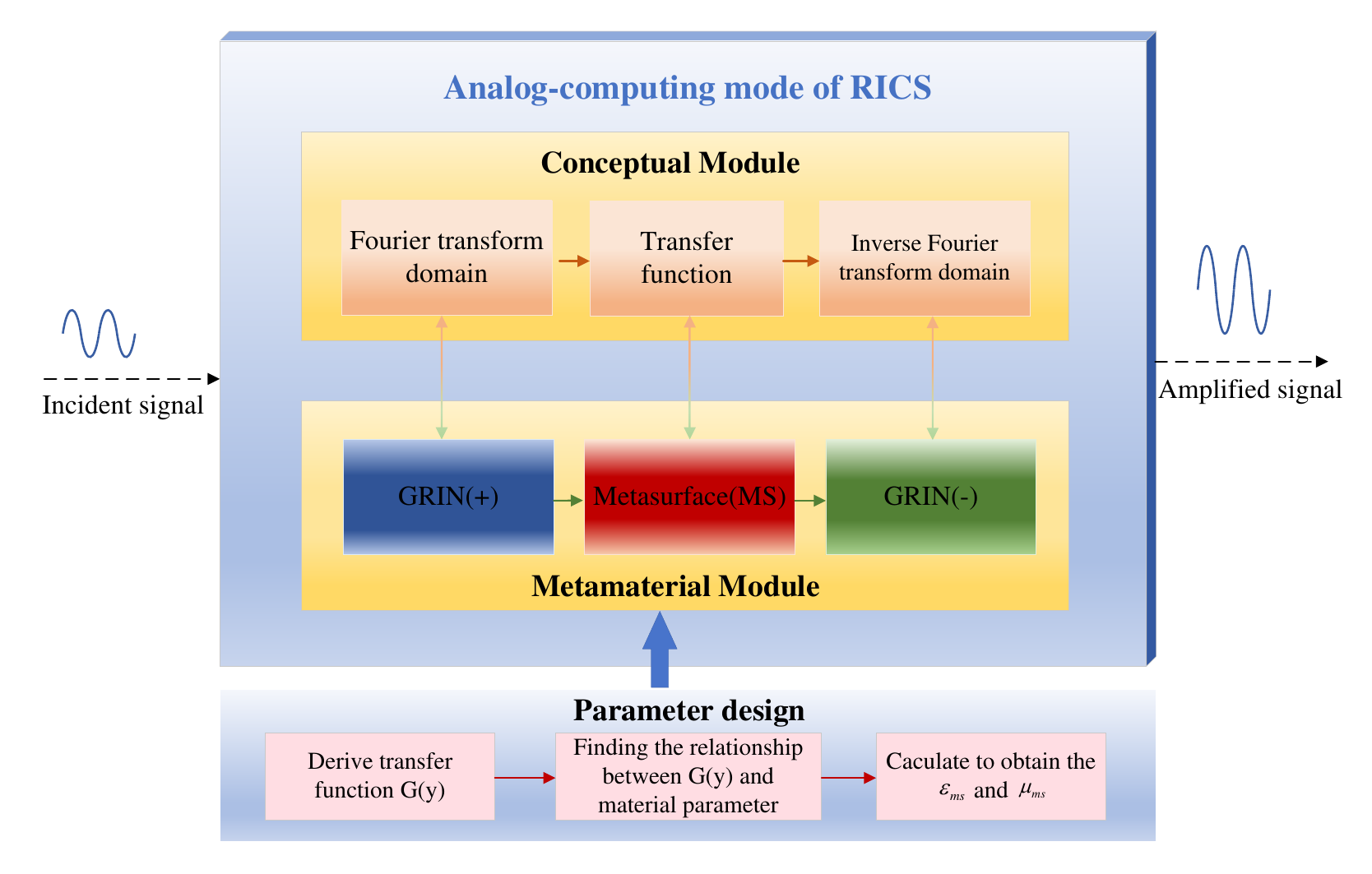}
    \caption{The design process of the analog computing mode of RICS.}
    \label{MS_design}
\end{figure}

For ease of understanding, the composite material design process corresponding to each element block of RICS analog computation mode is summarized in Fig.~\ref{MS_design}. Considering the constraints on the lateral dimensions of the GRIN, we normalize it within the lateral limits of the material, so the desired transfer function becomes $G(y)\!=\! \frac{1}{\Psi_{l} \cdot {{y}_{0}}}$, where ${{y}_{0}}=W/2$. 

Subsequently, by introducing the attenuation term ${{e}^{jk\Delta }}$, the MS modulates the amplitude of the incident wave to achieve the desired transfer function, denoted as ${{e}^{jk\Delta }}\!=\! 2/\left( {{\Psi }_{l}}W \right)$. Here, $k$ represents the wave number ($k$-space), which corresponds to the spatial frequency of the wave. $\Delta \!=\! {\lambda}_o/3$ denotes the thickness of the MS. The exponential decay term can alter the wave's amplitude, meaning that precise control over $\Delta $ allows for adjusting the amplitude to meet the desired requirements of the transfer function. Let ${{k}_{0}}\!=\!2\pi /{{\lambda }_{o}}$ be the wave number in free space, then we have 
\begin{equation} \label{k-space}
\begin{split}
    k&={{k}_{0}}\sqrt{[{{\varepsilon }_{ms}}(y)/{{\varepsilon }_{0}}][{{\mu }_{ms}}(y)/{{\mu }_{0}}]}\\
    &={{k}_{0}}\sqrt{\varepsilon _{ms}^{2}(y)/\varepsilon _{0}^{2}}\\
    &={{k}_{0}}{{\varepsilon }_{ms}}(y)/{{\varepsilon }_{0}}.  
\end{split}
\end{equation}
%\begin{table}[t]
%	\centering
%	\caption{Parameters of the materials}
%	\label{parameters}
%	\small
%	\begin{tabular}{|c|c|}
		%\noalign{\global\arrayrulewidth=0.3mm}
%		\hline
%		\textbf{Parameter}  &\textbf{Value } \\%&\textbf{Parameter}  & \textbf{Value } \\
		%\noalign{\global\arrayrulewidth=0.1mm}
%		\hline
 %       The permittivity of free space ${{\varepsilon }_{0}}$ &$4\pi \times {{10}^{-7}}H/m$\\ \hline
  %      The permeability of free space ${{\mu }_{0}}$ &$8.85\times {{10}^{-12}}F/m$\\ \hline
 %       The characteristic distance of GRIN $L_{g}$ &35$\mu m$\\ \hline
 %       The permittivity of the GRIN material ${{\varepsilon }_{c}}$ &2.01${{\varepsilon }_{0}}$\\ \hline
 %       The wavelength in free space ${{\lambda }_{o}}$ &3$\mu m$ \\ \hline
 %       The width of GRIN $W$ &10${{\lambda }_{o}}$\\ \hline
  %      The wave number $k_0$ in free space &$2\pi /{{\lambda }_{o}}$\\ \hline
  %      The permittivity of the MS ${{\varepsilon }_{ms}}\left( y \right)$ &$\frac{i{{\varepsilon }_{0}}}{{{k}_{0}}\Delta }\left( \ln {{\Psi }_{l}}W-\ln 2 \right)$\\ \hline
  %      The permeability of the MS ${{\mu }_{ms}}\left( y \right)$ &$\frac{i{{\mu }_{0}}}{{{k}_{0}}\Delta }\left( \ln {{\Psi }_{l}}W-\ln 2 \right)$\\ \hline
   %     The wave number k-space &${{k}_{0}}{{\varepsilon }_{ms}}\left( y \right)/{{\varepsilon }_{0}}$\\ \hline
   %     The thickness of MS $\Delta $ &${{\lambda }_{o}}/3$\\ \hline
		 
		%\noalign{\global\arrayrulewidth=0.3mm}
%		\hline
%	\end{tabular}
%\end{table}	
For the given point $y$, it can be observed that ${{\varepsilon }_{ms}(y)}$ and ${{\mu }_{ms}}(y)$ represent the permittivity and permeability of the MS, respectively. Suppose that the impedance of the GRIN roughly matches the impedance of the negative refracted wave in free space, we can use ${{\varepsilon }_{ms}}\left( y \right)/{{\varepsilon }_{0}}\!=\!{{\mu }_{ms}}\left( y \right)/{{\varepsilon }_{0}}$ to reduce reflections from the MS. Therefore, we have 
\begin{equation}
{{e}^{ik\Delta }}={{e}^{j{{k}_{0}}{{\varepsilon }_{ms}(y)}\Delta /{{\varepsilon }_{0}}}}=\frac{2}{{\Psi}_l W}.
\end{equation}

 Through derivation, we can determine the signal adjusting MS material, which should meet the following conditions
\begin{equation} \label{amplify}
\frac{{{\varepsilon }_{ms}}(y)}{{{\varepsilon }_{0}}}=\frac{{{\mu }_{ms}}(y)}{{{\mu }_{0}}}=\frac{j}{{{k}_{0}}\Delta }\left(\ln {{\Psi}_l}W-\ln 2\right).
\end{equation}

In general, the amplitude adjustment factor is limited by the properties of the materials, and the specific dielectric losses can restrict the adjustment effect. Additionally, the size of the system that we constructed cannot be infinite, so during the process of wave propagation, it is necessary to ensure that its width is less than $W$. The range of the imaginary part of the permittivity for  AZO and Si is approximately $0 \sim  10^2$~\cite{AZO,Si}. According to (\ref{amplify}), we need to ensure that the adjustment factor does not increase infinitely beyond the dielectric loss behavior of the material. Therefore, taking all these factors into consideration, the adjustment factor should be a limited value, which should be set carefully to fulfill the purpose of interference elimination in this paper.

\section{RICS-aided Autonomous Vehicular Networks }
\label{sention_3}
In this section, we use RICS for assisting autonomous driving networks and model the communication network. Further, in order to enhance the security of the autonmous driving network, we establish and analyze the RICS-aided optimization problem.

\subsection{Network Scenario}
%We consider a cellular network where the locations of the base stations (BSs) and the vehicle users are modeled as two independent homogeneous PPPs. Aligning with literature, we consider line boolean model for the blockages, and we assume that a fraction of these blockages are coated with RISs. 
In the considered RICS-aided autonomous driving scenario, as shown in Fig.~\ref{system}, there exists one BS, $M$ cellular vehicles (CVs) that communicate with the BS through V2I links, $N$ V2V communication pairs, and one RICS is implemented with uniform linear array (ULA). The CVs can offload the captured images to the MEC server for processing via V2I links, which can sometimes be shared by the V2V links for transmitting safety information. The set of CVs and V2V pairs is denoted by
${\cal M}=\{1, 2, \ldots, M\}$ and ${\cal N}=\{1, 2, \ldots, N\}$, respectively. Let a binary variable, $\alpha_{m,n}$, indicate the channel sharing result between the $m$th CV and the $n$th V2V pair. In this case, the CVs will cause interferences to V2V pairs if $\alpha_{m,n}=1$, and vice versa.

%To make full use of the available computation resources of the MEC server, current recommendations foresee that V2I communications are assigned the radio spectrum in an TDMA manner, while V2V communications reuse the radio spectrum of V2I communications.
 %To make full use of the available radio spectrum resources, current recommendations foresee that V2I communications are assigned the radio spectrum in an orthogonal manner, while V2V communications reuse the radio spectrum of V2I communications. 

\subsection{Autonomous Driving Model}
In the considered RICS-aided vehicular network, the CVs capture video sequences with the embedded camera sensors and need to infer the driving environment (e.g., the pedestrians ahead) in near-real-time. In this paper, the autonomous driving task (ADT) is defined as the task for object detection according to the captured images, which are processed by a pre-trained deep neural network (DNN) model deployed at the CVs. For the $i$th CV, an ADT can be characterized by a three-tuple of parameters, i.e., ${\cal T}_i(s_i, c_i, \sigma_i)$. Specifically, $s_i$ [bits] denotes the size of computation input data, $c_i$ [cycles] denotes the total number of CPU cycles required to accomplish the computation of $s_i$, and $\sigma_i$ [secs] denotes the maximum tolerable delay. % Specifically, if the wireless link is available, then the UAVs can offload the DL tasks to the MES and can also receive the results from the MES via the wireless link. Otherwise,  the DL tasks cannot be offloaded to the MES due to the wireless channel between UAVs and the MES is unavailable (e.g., wireless channel suffers deep fading). 

Without loss of generality, we assume that the DNN model deployed at the vehicles and the BS have different computational capabilities. For the $m$th CV processing the image data with a given quality $Q$, the inference accuracy achieved by the vehicles will be no larger than that given by the BS, i.e., $A_m(Q) = \lambda A_B(Q)$, $0 \leq \lambda \leq 1$, $\forall m \in \cal M$. 

\subsection{RICS-Based Communication Model}
 Due to the environmental changes between the CVs and the BS, the wireless channel condition may vary frequently, which could lead to the unavailability of the wireless channel in some cases. To improve the wireless link quality, RICS-aided wireless communications become necessary. The deployed RICS is equipped with $L$ reflecting elements that can be appropriately configured by a controller to reflect and process the impinging signals, thereby establishing line-of-sight (LoS) links between the BS and the vehicle users. We assume perfect knowledge of channel state information (CSI) is achieved at the BS, which can be fed back to the controller. Deploying the proposed RICS with RR+AC mode allows mathematical operations on the refracted signals that propagate through the intelligence computation layer. Therefore, the phase-shifted refracted signal can be adjusted to migrate the interference at the receiver of the $n$th V2V pair. 

In our considered RICS-aided autonomous vehicular wireless system, we denote the channels from the desired $m$th CV to the RICS, from the RICS to the BS, %from the sender of the $n$th V2V pair to the RICS, 
and from the RICS to the receiver of the $n$th V2V pair, are ${\textbf h}_{m,R} \in \mathbb{C}^{L \times 1}$, ${\textbf h}_{R,B} \in \mathbb{C}^{1 \times L}$, %$\textit{\textbf h}_{s,R}^n \in \mathbb{C}^{L \times 1}$, 
and ${\textbf h}_{R,n} \in \mathbb{C}^{L \times 1}$, respectively.
Otherwise, we define path loss as $P_L=\sqrt{{{C}_{0}}{{d}^{-\alpha }}}$, where $C_{0}$ represents the path loss at the reference distance of $d_{0}=1$, $d$ is the distance between specific links, an $\alpha$ is the path loss exponent.
In the considered system, each of the $L$ CV$_m$-RICS-Rx$_n$ links is modeled as a Rician channel in order to take into account the LoS contribution and the non-LOS (NLoS) multipath components. Therefore, the channel gain %from the $m$th CV to the $k$th antenna of the BS, and the channel gain from the $k$th antenna of the BS to the Rx of the $n$th V2V pair via the $l$th reconfigurable element of the RICS 
is given by
\begin{equation}\label{h_mr}
{\textbf h}_{m,R}\!=\!  P_L \left(\sqrt{\frac{\kappa_{m,R}}{1\!+\!\kappa_{m,R}}}{\textbf h}_{m,R}^{\rm LoS}+\sqrt{\frac{1}{1\!+\!\kappa_{m,R}}}{\textbf h}_{m,R}^{\rm NLoS}\right),
\end{equation}
\begin{equation}\label{h_rn}
{\textbf h}_{R,n}\!=\!  P_L \left(\sqrt{\frac{\kappa_{R,n}}{1\!+\!\kappa_{R,n}}}{\textbf h}_{R,n}^{\rm LoS}+\sqrt{\frac{1}{1\!+\!\kappa_{R,n}}}{\textbf h}_{R,n}^{\rm NLoS}\right),
\end{equation}
where $\kappa_{m,R}$ and $\kappa_{R,n}$ denote the Rician factor related to small-scale fading. Moreover, the LoS components, ${\textbf h}_{m,R}^{\rm LoS}$ and ${\textbf h}_{R,n}^{\rm LoS}$, consist of the ULA array response, and each element of the NLoS component, ${\textbf h}_{m,R}^{\rm NLoS}$ and ${\textbf h}_{R,n}^{\rm NLoS}$, follow an i.i.d. complex Gaussian distribution with zero mean and unit variance.

 The channel gains of the direct links from the $m$th CV to the BS and from the sender to the receiver of the $n$th V2V pair are denoted by ${h}_{m,B}$ and $h_{n}$, respectively. Moreover, the interference channel gain from the $m$th CV to the receiver of the $n$th V2V pair, from the sender of the $n$th V2V pair to the BS, is given as $h_{m, n}$ and ${h}_{n, B} $, respectively. These channels are assumed to be perfectly estimated and quasi-static, hence remaining nearly constant during the transmissions.

%To migrate the interference at the receiver of the $n$th V2V pair, we deploy the proposed RICS with RR+AC mode.
We denote $P_m$ and $P_t$ as the transmission power of the $m$th CV and the Tx of the $n$th V2V pair, respectively. $W$ stands for the transmission bandwidth between the CVs and the BS. $\xi_0$ indicates the noise power spectral density. So the received SINR of the $m$th CV at the BS and the received SINR of the $n$th V2V pair can be obtained as
\begin{equation} \label{sinr_m}
\gamma_B^m = \frac{{P_m} \left |{h}_{m,B} +  {\textbf h}_{R,B} \mathbf{\Phi}_r {\textbf h}_{m,R}  \right |^2 }{\sum_{n=1}^{N}{\alpha_{m,n} P_t} \left |{h}_{n,B}   \right |^2 + W \xi_0},
\end{equation}
%\gamma_m\!=\! \frac{{P_m} \left |h_{m,B}+  \textit{\textbf h}_{R,B} \mathbf{\Phi} \textit{\textbf h}_{m,R}  \right |^2 }{\sum_{n=1}^{N}{\alpha_{m,n} P_t} \left |h_{n} \!+\! \sum_{k=1}^{K} \textit{\textbf h}_{k,b} \mathbf{\Phi}_k \textit{\textbf h}_{n,k}  \right |^2 \!+\! \sigma^2}
\begin{equation} \label{sinr_n}
\gamma_n= \frac{{P_t} \left |h_{n} \right |^2 }{\sum_{m=1}^{M}{\alpha_{m,n} P_m}  \big|h_{m,n} + {\textbf h}_{R,n}^H \mathbf{\Phi}_t {\textbf h}_{m,R}\big|^2  + W\xi_0}.
\end{equation}

%where $P_m$ and $P_t$ denote the transmission power of the $m$th CV and the Tx of the $n$th V2V pair, respectively. %$\epsilon$ indicates the power ratio of the refracted signal to the reflected signal, so $P(\epsilon)$ denotes the refracted offset power for migrating the interference at the receiver of the $n$th V2V pair. 
%$W$ stands for the transmission bandwidth between the CVs and the BS. $\xi_0$ indicates the noise power spectral density. 
Note that the refracted channel ${\textbf h}_{R,n}^H \mathbf{\Phi}_t {\textbf h}_{m,R}$ is in the opposite phase with the direct channel $h_{m,n}$, thereby eliminating the interfering signal at the receiver of the $n$th V2V pair.
%\gamma_n\!=\! \frac{{P_t} \left |h_{n}+ \sum_{k=1}^{K}  \textit{\textbf h}_{k,r} \mathbf{\Phi}_k \textit{\textbf h}_{s,k}  \right |^2 }{\sum_{m=1}^{M}{\alpha_{m,n} P_m} \left |h_{m} \!+\! \sum_{k=1}^{K} \textit{\textbf h}_{k,r} \mathbf{\Phi}_k \textit{\textbf h}_{m,k}  \right |^2 \!+\! \sigma^2},
Accordingly, the achievable uplink rate of the $m$th CV and the achievable data rate of the $n$th V2V pair is given by 
\begin{equation}\label{R_m}
R_B^m=W {\rm log}_2{\left ( 1+ \gamma_B^m \right )}, \ \ \forall m \in {\cal M},
\end{equation}
\begin{equation}\label{R_n}
R_n=W {\rm log}_2{\left ( 1+ \gamma_n \right )}, \ \forall n \in {\cal N}.
\end{equation}
%where $W$ stands for the transmission bandwidth between the CVs and the BS, which can be further divided into $M$ orthogonal sub-bands for the offloading communication.

\subsection{RICS-Based Computation Model}
Without loss of generality, the partial offloading model is investigated, which is more general in practice since it can fully utilize the computation resources in both the CVs and the BS. Specifically, for the $m$th CV, the offloading ratio, $\rho_m$, is defined as the ratio (or portion) of the DL tasks that are offloaded to the BS. Accordingly, $1-\rho_m$ indicates the ratio of data to be processed locally. Suppose that the time-interdependency between each video frame within the DL task is ignored, then ${\cal J}_i$ can be divided into two parts, i.e., $\rho_m s_m$ (bits) is offloaded to the BS while $(1-\rho_m)s_m$ (bits) is processed locally at the CVs.
%In the following, we first formulate the latency-minimization problem as a piecewise-convex problem and then derive the optimal offloading ratio (i.e., $\beta_i^*$). Then, two cases are considered where we analyze $\beta_i^*$ to minimize energy consumption and total cost, respectively. Finally, a special scenario is considered assuming that the UAV can distribute the ML tasks according to the data quality $\cal Q$, and the optimal tasks segmentation ratio is derived accordingly.

For the local computing, $(1-\rho_m) s_m$ of ${\cal J}_m$ is processed locally at the CV.  %Considering that the more DL lower-layers the longer time processing the tasks at the UAV, denote $f^i_{l}$ as the CPU-cycle frequency (i.e., CPU cycles per second) of the $i$-th UAV, $i \in {\cal N}$, 
The local computation delay of the $m$th CV is  calculated as 
\begin{equation}\label{partial_local_delay}
\tau^m_{l} = \left ( 1-\rho_m \right )  \frac{c_m}{f_m},
\end{equation}
where $f_m$ indicates the computing resource of the $m$th CV.

For the offloading computing,  $\rho_m s_m$ of ${\cal J}_m$ is offloaded to the BS. The total delay introduced by the offloading computing is given by
 \begin{equation}\label{partial_delay_offloading}
 \tau^m_{o} = \rho_m \left (  \frac{s_m}{R_B^m}+\frac{c_m}{F} \right ),
\end{equation}
where $F$ denotes the computation resource of the BS.

So the total delay of the $m$th CV introduced by the partial offloading scheme is calculated as
\begin{equation}\label{partial_overall_delay}
\tau_m={\rm max} \{\tau^m_{l},  \tau^m_{o}  \}.
\end{equation}

Accordingly, the average inference accuracy is obtained as 
\begin{equation}\label{Accuracy}
\tilde{A}_m(Q)= \left ( 1- \rho_m \right )A_m(Q) + \rho_m A_B(Q),
\end{equation}
where $A_m(Q)$ and $A_B(Q)$ denote the inference accuracy given by the shallow CNN at the $m$th CV and the deep CNN at the BS,  respectively.

\subsection{Problem Formulation and Analysis}
\label{sention_4}
\subsubsection{Problem Formulation}
We define the driving safety coefficient as 
\begin{equation}\label{saftey}
{S}_m = \frac{\tilde{A}_m(Q)}{\tau_m} = \frac{\left ( 1- \rho_m \right )A_m(Q) + \rho_m A_B(Q)}{{\rm max} \{\tau^m_{l},  \tau^m_{o}  \}} .
\end{equation}

By substituting $A_m(Q) = \lambda A_B(Q)$ into (\ref{saftey}), we have
\begin{equation}\label{saftey1}
{S}_m = \frac{A_B(Q) \left( \lambda +  \rho_m \left ( 1- \lambda  \right ) \right )}{{\rm max} \{\tau^m_{l},  \tau^m_{o}  \}} .
\end{equation}

In this paper, we aim to maximize the safety coefficient of the CVs while satisfying the outage probability of V2V pairs:\hspace{-2mm}
\begin{subequations} \label{binary_original_problem}
\begin{align}
& \;\;\;\; \mathbb{P}: \ \underset{\bm{\alpha}, \mathbf{\Phi}_x, \bm{\rho}}{\rm max}\;\;  \sum\limits_{m\in \mathcal{M}}S_m  \notag \\
& \;\;\;\;\;\;{{s}}{{.t}}{\rm{.}}\;\;\; \; {\rm Pr}\{\gamma_n \leq \gamma_{th}\} \leq P_{outage}, \\
& \;\;\;\;\;\;\;\;\;\;\;\;\;\; \;  \alpha_{m,n} \in \{0,1\}, \ \forall m \in {\cal M}, \forall n \in {\cal N},  \\
& \;\;\;\;\;\;\;\;\;\;\;\;\;\;  \sum_{n=1}^{N} \alpha_{m,n} \le 1, \ \forall m \in {\cal M}, \forall n \in {\cal N},  \\
& \;\;\;\;\;\;\;\;\;\;\;\;\;\;\left| {{\theta }_{t}}-{{\theta }_{r}} \right|=\frac{\pi }{2}\,\,or\,\,\frac{3\pi }{2},1\le l\le L  \\
& \;\;\;\;\;\;\;\;\;\;\;\;\;\; \; \beta_l^{r}+\beta_l^{t}=1, \ 1 \le l \le L, \\
& \;\;\;\;\;\;\;\;\;\;\;\;\;\; \; \rho_m \in [0,1], \ \forall m \in {\cal M},
 \end{align}
\end{subequations}
where $\bm \alpha = \{\alpha_{m,n}, \forall m, n\}$,  $\bm \rho=\{\rho_1, \rho_2, \ldots, \rho_m\}$.
%& \;\;\;\;\;\;\;\;\;\;\;\;\;\;\mathbf{C5}: \; 0 \le f_{m} \le F, \ \forall m \in {\cal M}, \\ & \;\;\;\;\;\;\;\;\;\;\;\;\;\;\mathbf{C6}: \; \sum_{m=1}^{M} f_{m} \le F,$\bm f=\{f_1, f_2, ..., f_m\}$.$\bm{\Theta}=\{\bm \theta_1, \bm \theta_2, ..., \bm \theta_k\}$,

\subsubsection{Outage Probability Analysis of V2V Pairs}
In this part, we analyze the outage probability of V2V pairs, which should meet the constraint (\ref{binary_original_problem}a). Specifically, the outage probability constraint in (\ref{binary_original_problem}a) can
be rewritten as 
${\rm Pr}\{\gamma_n \leq \gamma_{th}\} = \mathbb{E} \left [u \left(\gamma_{th}- \gamma_n\right)  \right ]$, where $\mathbb{E}[x]$ denotes $x$'s expected value and $u(x)$ is a step function. Since there exist many smooth approximations of the step function, so we let $\hat{u}_\omega(x)=\frac{1}{1+e^{-\omega x}}$ denote a smooth approximation of the step function $u(x)$ with a smooth parameter $\omega$ for controlling the approximation error.
 
 By replacing the step $u(x)$ with its smooth approximation $\hat{u}_\omega(x)$, we can obtain an approximation of the constraint (\ref{binary_original_problem}a):
\begin{equation} \label{proof01}
\mathbb{E} \left [\hat{u}_\omega \left(\gamma_{th}- \gamma_n\right)  \right ] \leq P_{outage}.
\end{equation} 
 
According to Jensen’s inequality, the left-hand side (LHS) of (\ref{proof01}) can be
upper bounded by
\begin{equation} \label{proof03}
 \begin{aligned}
\mathbb{E} \left [\hat{u}_\omega \left(\gamma_{th}- \gamma_n \right)  \right ] &\leq  \hat{u}_\omega \left(\mathbb{E} \left[\gamma_{th}- \gamma_n \right ] \right)  \\
&= \hat{u}_\omega \left( \gamma_{th}- \mathbb{E} \left[ \gamma_n \right ] \right).
\end{aligned}
\end{equation}

As for $\mathbb{E} \left[ \gamma_n \right ]$, by substituting (\ref{sinr_n}) into (\ref{proof03}), we have
\begin{equation} \label{proof04}
 \begin{aligned}
\mathbb{E} \left[ \gamma_n \right ] &= \mathbb{E} \left[\frac{{P_t} \left |h_{n} \right |^2 }{\sum_{m=1}^{M}{\alpha_{m,n} P_m}  \big|h_{m,n} \!+\! {\textbf h}_{R,n}^H \mathbf{\Phi}_t {\textbf h}_{m,R}\big|^2  \!+\! W\xi_0}  \right ] \\
\!&=\! \frac{\mathbb{E} \left[ {P_t} \left |h_{n} \right |^2 \right ] }{\mathbb{E} \left[\sum_{m=1}^{M}{\alpha_{m,n} P_m}  \big|h_{m,n} \!+\! {\textbf h}_{R,n}^H \mathbf{\Phi}_t {\textbf h}_{m,R}\big|^2 \!+\! W\xi_0\right ]} \\
\!&=\! \frac{{P_t} \cdot \mathbb{E} \left[ \left |h_{n} \right |^2 \right ] }{\sum_{m=1}^{M}{\alpha_{m,n} P_m} \cdot \mathbb{E} \left[\big|h_{m,n} \!+\! {\textbf h}_{R,n}^H \mathbf{\Phi}_t {\textbf h}_{m,R}\big|^2\right ] \!+\! W\xi_0}.
\end{aligned}
\end{equation}

\begin{figure*}[!h]
	\begin{equation}
		\begin{aligned}\label{proof05}
		 &\mathbb{E} \left[\big|h_{m,n} \!+\! {\textbf h}_{R,n}^H \mathbf{\Phi}_t {\textbf h}_{m,R}\big|^2\right ]\\
		&=\mathbb{E} \left[\left|h_{m,n}\right|^2\right ] \!+\! \mathbb{E} \left[\left|\left(\sqrt{\frac{\kappa_{R,n}}{1\!+\!\kappa_{R,n}}}{\textbf h}_{R,n}^{\rm LoS}+\sqrt{\frac{1}{1\!+\!\kappa_{R,n}}}{\textbf h}_{R,n}^{\rm NLoS} \right)^H \mathbf{\Phi}_t \left(\sqrt{\frac{\kappa_{m,R}}{1\!+\!\kappa_{m,R}}}{\textbf h}_{m,R}^{\rm LoS}+\sqrt{\frac{1}{1\!+\!\kappa_{m,R}}}{\textbf h}_{m,R}^{\rm NLoS} \right)\right|^2\right ] \\
		&=\mathbb{E} \left[\left|h_{m,n}\right|^2\right ] \!+\! \mathbb{E} \left[\left| \sqrt{\frac{1}{\left(1\!+\!\kappa_{R,n}\right) \left(1\!+\!\kappa_{m,R} \right)}} \right|^2 \cdot \left|{\textbf H}_1 +{\textbf H}_2+{\textbf H}_3+{\textbf H}_4 \right|^2 \right ] \\ 
		&=\mathbb{E} \left[\left|h_{m,n}\right|^2\right ] \!+\! \frac{1}{\left(1\!+\!\kappa_{R,n}\right) \left(1\!+\!\kappa_{m,R} \right)} \left(\mathbb{E}\left[\left|{\textbf H}_1\right|^2 \right ]+\mathbb{E}\left[\left|{\textbf H}_2\right|^2 \right ] +\mathbb{E}\left[\left|{\textbf H}_3\right|^2 \right ] +\mathbb{E}\left[\left|{\textbf H}_4\right|^2 \right ]\right)\\
		&= 1\!+\! \frac{1}{\left(1\!+\!\kappa_{R,n}\right) \left(1\!+\!\kappa_{m,R} \right)} \left(\left|\sqrt{\kappa_{R,n} \kappa_{m,R}} \cdot {\textbf h}_{R,n}^{\rm LoS} \mathbf{\Phi}_t {\textbf h}_{m,R}^{\rm LoS}\right|^2 +N \kappa_{R,n} +N \kappa_{m,R} +N\right).
		\end{aligned}
	\end{equation}
	\rule{\textwidth}{1pt}
\end{figure*}

\begin{figure*}[!h]
	\begin{equation}
    \begin{aligned}\label{proof06}
     &\mathbb{E} \left[\big|h_{m,n} \!+\! {\textbf h}_{R,n}^H \mathbf{\Phi}_t {\textbf h}_{m,R}\big|^2\right ]\\
     & = 1 \!+\! \frac{1}{\left(1\!+\!\kappa_{R,n}\right) \left(1\!+\!\kappa_{m,R} \right)} \left(\left|{\textbf H}_1\right|^2 +N \kappa_{R,n} +N \kappa_{m,R} +N\right) \\
     &= 1\!+\! \frac{1}{\left(1\!+\!\kappa_{R,n}\right) \left(1\!+\!\kappa_{m,R} \right)} \left(\left|\sqrt{\kappa_{R,n} \kappa_{m,R}} \cdot {\textbf h}_{R,n}^{\rm LoS} \mathbf{\Phi}_t {\textbf h}_{m,R}^{\rm LoS}\right|^2 +N \kappa_{R,n} +N \kappa_{m,R} +N\right).
    \end{aligned} 
    \end{equation}
	\rule{\textwidth}{1pt}
\end{figure*}

\begin{figure*}[!h]
	\begin{equation}
		\begin{aligned}\label{proof07}
		\mathbb{E} \left[ \gamma_n \right ] &= \frac{{P_t} \cdot \mathbb{E} \left[ \left |h_{n} \right |^2 \right ] }{\sum_{m=1}^{M}{\alpha_{m,n} P_m} \cdot \mathbb{E} \left[\big|h_{m,n} \!+\! {\textbf h}_{R,n}^H \mathbf{\Phi}_t {\textbf h}_{m,R}\big|^2\right ] \!+\! W\xi_0} \\
    &=\frac{{P_t} \cdot \mathbb{E} \left[ \left |h_{n} \right |^2 \right ] }{\sum_{m=1}^{M}{\alpha_{m,n} P_m} \left(1\!+\! \frac{1}{\left(1\!+\!\kappa_{R,n}\right) \left(1\!+\!\kappa_{m,R} \right)} \left(\left|\sqrt{\kappa_{R,n} \kappa_{m,R}} \cdot {\textbf h}_{R,n}^{\rm LoS} \mathbf{\Phi}_t {\textbf h}_{m,R}^{\rm LoS}\right|^2 \!+\! N \kappa_{R,n} +N \kappa_{m,R} \!+\! N\right) \right) \!+\! W\xi_0}.
		\end{aligned} 
	\end{equation}
	\rule{\textwidth}{1pt}
\end{figure*}

Then, we focus our attention on the computation of $\mathbb{E} \left[\big|h_{m,n} \!+\! {\textbf h}_{R,n}^H \mathbf{\Phi}_t {\textbf h}_{m,R}\big|^2\right ]$ in (\ref{proof04}). It is assumed that $h_{m,n}$,  ${\textbf h}_{R,n}$ and ${\textbf h}_{m,R}$ are independent of each other, so we can derive (\ref{proof05})-(\ref{proof07}), %shown at the bottom of the previous page, 
where 
\begin{subequations}
\begin{align}
& {\textbf H}_1 \triangleq \sqrt{\kappa_{R,n} \kappa_{m,R}} \cdot {\textbf h}_{R,n}^{\rm LoS} \mathbf{\Phi}_t {\textbf h}_{m,R}^{\rm LoS}, \\
& {\textbf H}_2 \triangleq \sqrt{\kappa_{R,n}} \cdot {\textbf h}_{R,n}^{\rm LoS} \mathbf{\Phi}_t {\textbf h}_{m,R}^{\rm NLoS}, \\
& {\textbf H}_3 \triangleq \sqrt{\kappa_{m,R}} \cdot {\textbf h}_{R,n}^{\rm NLoS} \mathbf{\Phi}_t {\textbf h}_{m,R}^{\rm LoS}, \\
& {\textbf H}_4 \triangleq  {\textbf h}_{R,n}^{\rm NLoS} \mathbf{\Phi}_t {\textbf h}_{m,R}^{\rm NLoS}.
\end{align}
\end{subequations}

As for the direct V2V link with $h_{m,n}\sim {\cal C N}(0,1) $ and the desired cascaded channels, the equations $\mathbb{E} \left[\left|h_{m,n}\right|^2\right ] \!=\! 1$, $\mathbb{E}\left[\left|{\textbf H}_1\right|^2 \right ]\!=\!\left|{\textbf H}_1\right|^2$, $\mathbb{E}\left[\left|{\textbf H}_2\right|^2 \right ]\!=\!N \kappa_{R,n}$, $\mathbb{E}\left[\left|{\textbf H}_3\right|^2 \right ]\!=\!N  \kappa_{m,R}$, and $\mathbb{E}\left[\left|{\textbf H}_4\right|^2 \right ]\!=\!N$ hold. By substituting (\ref{proof05}) into (\ref{proof03}), we have 
\begin{equation} \label{proof08}
 \begin{aligned}
\mathbb{E} \left [\hat{u}_\omega \left(\gamma_{th}- \gamma_n \right)  \right ] &\approx \hat{u}_\omega \left( \gamma_{th}- \mathbb{E} \left[ \gamma_n \right ] \right) \\
&=\hat{u}_\omega \left( \gamma_{th}-\tilde{\gamma}_{n}(\bm{\alpha}, \mathbf{\Phi}_x) \right),
\end{aligned}
\end{equation}

where $\tilde{\gamma}_n (\bm{\alpha}, \mathbf{\Phi}_x) = \mathbb{E} \left [ \gamma_n \right ]$ with $\mathbb{E} \left[ \left |h_{n} \right |^2 \right ]=1$.

According to (\ref{proof01}) and (\ref{proof08}), the constraint (\ref{binary_original_problem}a) can be rewritten as 
\begin{equation}\label{proof09}
\hat{u}_\omega \left( \gamma_{th}-\tilde{\gamma}_{n}(\bm{\alpha}, \mathbf{\Phi}_x) \right) \leq P_{outage}.
\end{equation}

Since we have $\hat{u}_\omega(x)=\frac{1}{1+e^{-\omega x}}$, by performing the inequality transformation, (\ref{proof09}) can be rewritten as 
\begin{equation} \label{proof02}
\tilde{\gamma}_n (\bm{\alpha}, \mathbf{\Phi}_x) \geq \gamma_{th}+\frac{1}{\omega} {\rm ln} \left ( \frac{1}{P_{outage}}-1 \right )\triangleq \tilde{\gamma}_c.
\end{equation}

Based on (\ref{proof02}), the constraint (\ref{binary_original_problem}a) can be conveniently rewritten to tackle the optimization problem $\mathbb{P}$. Meanwhile, to make the original optimization problem $\mathbb{P}$ more tractable,
we first relax the binary variables in (\ref{binary_original_problem}b) into continuous variables, which yields the following equivalent problem:
\begin{subequations} \label{equivalent_problem}
\begin{align}
& \;\;\;\; \tilde{\mathbb{P}}: \ \underset{\bm{\alpha}, \mathbf{\Phi}_x, \bm{\rho}}{\rm max}\;\;  \sum\limits_{m\in \mathcal{M}} S_m  \notag \\
& \;\;\;\;\;\;{{s}}{{.t}}{\rm{.}}\;\;\; \; 0 \leq \alpha_{m,n} \leq 1, \ \forall m \in {\cal M}, \forall n \in {\cal N},  \\
& \;\;\;\;\;\;\;\;\;\;\;\;\;\;  (\ref{binary_original_problem}c) - (\ref{binary_original_problem}f), (\ref{proof02}). \notag
\end{align}
\end{subequations}

By using linear relaxation, the objective value of the equivalent problem $\tilde{\mathbb{P}}$ usually provides an upper bound for the objective value of the original problem $\mathbb{P}$.  Note that in the equivalent problem $\tilde{\mathbb{P}}$, there are three optimization
variables, i.e., the spectrum sharing strategy for V2V links $\bm{\alpha}$, the RICS reflection-coefficient matrix $\mathbf{\Phi}_x$, and the offloading strategy $\bm{\rho}$. 

However, the problem $\mathbb{P}$ is non-convex and thus it is in general difficult and prohibitive to find the globally optimal solution due to the coupling of the variables.
%since the left-hand-side (LHS) of (\ref{proof02}) is not concave.  
%Due to the coupling of the variables, it is, in general, difficult and prohibitive to find the globally optimal solution for this non-convex problem. 
Motivated by this, in the following, we develop efficient algorithms to find high-quality solutions for such a challenging problem.

\section{Proposed Driving Safety Design}\label{sention_5}

According to the analysis in Section~\ref{sention_3}, the original problem can be rewritten as the equivalent problem $\tilde{\mathbb{P}}$. Due to the non-convexity of the optimization problem as well as the coupling between the three variables, it is difficult to obtain the global optimal solution directly. Therefore, we consider splitting the problem $\tilde{\mathbb{P}}$ into three subproblems and using the alternating optimization method. This method works by updating only one variable at a time in each iteration while keeping the other variables fixed. By iteratively updating all variables in a loop until a convergence condition is satisfied.

\subsection{Optimization of the Task Offloading Ratio}
Since the problem $\tilde{\mathbb{P}}$ is a fractional programming (FP) problem and the variable $\bm{\rho}$ to be optimized appears only in the objective function, so for the fixed channel sharing policy $\bm{\alpha}$ and the reflection-refraction coefficient matrix ${\mathbf{\Phi}}_x$, we apply the quadratic transform algorithm \cite{FP_Quad}, to convert the original optimization problem into a linearized form. So we introduce a coefficient, $\bm{\mu}$, to reformulate the programming problem as
\begin{subequations} \label{Quad}
\begin{align}
& \text{        }\underset{\{\bm{\rho} ,\bm{\mu} \}}{\mathop{\max }}\,\sum\limits_{m\in \mathcal{M}}{\left(2\mu \sqrt{(1-\lambda ){{A}_{B}}(Q){\bm{\rho }}+\lambda {{A}_{B}}(Q)}-{{\mu }^{2}}{{\tau }}(\bm{\rho })\right)} \nonumber \\ 
%& \text{   }{{s}}{{.t}}{\rm{.}}\;\;\;\;\;\;\;\;\;\;\;\;\;\;\;\;\;\;\;\;\;\;\;\;\;\;\;\;\;\; \bm{\mu} \in \mathbb{R}\\ 
&{{s}}{{.t}}{\rm{.}}\;\;\;\;\;(\ref{binary_original_problem}f).\nonumber
\end{align}
\end{subequations}

\begin{algorithm}[t]
\caption{Quadratic Transformation algorithm}\label{a1}
\begin{algorithmic}
%\small
\STATE 
\STATE Initialize the feasible solutions $\bm{\rho}$ and the optimal target vector $\bm{S}^{0}$, the maximum number of iteration times $k_{max}$, and let $k=1$;
\WHILE{$k \leq k_{max}$}
    \STATE update $\bm{\mu} =\frac{\sqrt{(1-\lambda ){{A}_{B}}(Q)(\bm{\rho} ')+\lambda {{A}_{B}}(Q)}}{{{\tau }}(\bm{\rho} ')}$
    \STATE Solving the problem (\ref{Quad}) to obtain $\bm{\rho}^{*}$ and $\bm{S}^{k}$;
    \IF{$\sum\bm{S}^{k}-\sum\bm{S}^{k-1} \leq \delta$}
        \STATE break;
    \ENDIF
\ENDWHILE
\STATE \textbf{return} $\bm{\rho}_{opt}=\bm{\rho}^*$, $\bm{S}_{opt}=\bm{S}^{k}$.
\end{algorithmic}
\label{a1}
\end{algorithm}

%If and only if $\bm{\rho}^*$ maximizes the original scale problem, 
There exists a paring of $\bm{\rho}^*$ and some $\bm{\mu}^*$ that maximizes the original scale problem. By defining $\bm{\mu}^* \!=\! \arg\max f(\bm{\rho}, \bm{\mu})$, then we establish that $f(\bm{\rho},\bm{\mu}^*) \!=\! \bm{S}$. Since $f(\bm{\rho}, \bm{\mu})$ exhibits concavity under the fixed $\bm{\rho}$ and $\bm{\mu}$, as a result, we can perform a convex optimization on $\bm{\mu}$. By employing the quadratic transformation algorithm, as shown in \textbf{Algorithm}~\ref{a1}, $\bm{\mu}$ and $\bm{\rho}$ can be alternately and iteratively updated and eventually leading to the convergence of the fractional planning problem and the attainment of the global optimal solution.

After completing the transformation of the above optimization problem, we iteratively update using the auxiliary variable $\bm{\mu}$, specifically expressed below
\begin{equation} \label{update}
\bm{\mu} [t+1]=\frac{\sqrt{A'({\bm{\rho }}[t])}}{B'({\bm{\rho }}[t])},
\end{equation}
where $t$ denotes the iterative subscript, the function ${A'}$ represents the function of the molecular part of $\bm{S}$ about $\bm{\rho}$, while ${B'}$ pertains to the denominator part of $\bm{ S}$ with $\bm{\rho}$. 
%To solve this problem, we propose the quadratic transformation algorithm, as illustrated in \textbf{Algorithm}~\ref{a1}. By alternating updating $\bm{\mu}$ iteratively and solving for $\bm{\rho}$ in the transformed poly-ratio problem, the convergence can be guaranteed. 
\hspace{-2mm}
\subsection{Optimization of the Spectrum Sharing Strategy}
Given ${\mathbf{\Phi}}_x$ and $\bm{\rho}$, we perform alternating optimization of spectrum sharing strategy $\bm{\alpha}$, so $\tilde{\mathbb{P}}$ in (\ref{equivalent_problem}) can be rewritten as\hspace{-2mm}
\begin{subequations} \label{reformulate}
\begin{align} 
& \tilde{\mathbb{P}}:\;\;\;\;\;\;\underset{\{\bm{\alpha}\}}{\mathop{\max }}\,\text{  }\sum\limits_{m\in \mathcal{M}}{{{S}_{m}}} \nonumber \\ 
&{{s}}{{.t}}{\rm{.}}%\text{     0}\le {\bm{\alpha }_{m,n}}\le 1\text{ },\ \forall m \in {\cal M}, \forall n \in {\cal N}, \\ 
%& \text{       }\;\;\;\;\;\sum\limits_{n\in \mathcal{N}}{{\bm{\alpha }_{m,n}}\le 1},\ \forall m \in {\cal M}, \forall n \in {\cal N}, \\ 
 \text{       }\;\;\;\;\;{{\tilde{\mathop{\gamma }}\,}_{n}}(\bm{\alpha} )\ge {\tilde{\gamma }_{c}} \\
&\;\;\;\;\;\;\;\;\;\;\;\; (\ref{binary_original_problem}c),    (\ref{equivalent_problem}a). \nonumber
\end{align}
\end{subequations}

By observing (\ref{binary_original_problem}), it is evident that the molecule remains in a stationary state when the optimization variable $\bm{\rho}$ is fixed. For the total denominator delay $\tau_m$, since the local computation delay is not influenced by the optimization variable $\bm{\alpha}$, so the first component of the max function remains constant. The function ${R_{B}^{m}}$  is the only component affected by $\bm{\alpha}$. By considering the properties of the harmonic average, the problem (\ref{reformulate}) can be reformulated as minimizing the total sum of delays $\sum\nolimits_{i=1}^{\mathcal{M}}{{{\tau }_{i}}}$. However, hindered by its non-convex nature, investigating advanced techniques to address this challenge becomes necessary.
 
Even though $\tau _{l}^{m}$ and ${\tau _{o}^{m}}$ are convex, the composite function formed by the $\rm max$ operation may exhibit non-convexity. Therefore, directly solving the above optimization problem might be quite challenging. To simplify the solving process, it is necessary to transform the problem into a different form that is more trackable. Following several classical convex functions~ \cite{log-sum-exp}, we can obtain 
$\max \{{{x}_{1}},{{x}_{2}},\ldots,{{x}_{n}}\}\le \log (\sum\limits_{i\in {N}}{{{e}^{{{x}_{i}}}}})$. By applying the log-sum-exp technique, the max operation is transformed into a smooth and differentiable approximate expression, which helps in utilizing convex optimization techniques to find the global optimal solution.
Therefore, by reformulating the original function in the form of log-sum-exponential, we are able to derive an upper bound for its optimization. This method enables us to effectively manage the non-convexity of ${\tau}_{i}$ and explore strategies for optimization within a more tractable framework, i.e., 
\begin{equation} \label{tau}
\underset{\{\bm{\alpha}\}}{\text{min }}\sum\limits_{m\in \mathcal{M}}{\tau _{m}^{ub}}=\sum\limits_{m\in \mathcal{M}}{\log \left( {{e}^{\tau _{l}^{m}}}+{{e}^{\tau _{o}^{m}}} \right)}.
\end{equation}

At this stage, our primary focus is to guarantee the convexity of $\tau^m_{o}$. Given that the inverse of a concave function is convex, our objective is to establish the concave nature of $R_{B}^{m}$. We first rewrite $R_{B}^{m}$ by employing the Difference of Concave (DC) functions
\begin{equation}
\begin{split}
R_{B}^{m}(\bm{\alpha})=\underbrace{W{{\log }_{2}}\left({{\Xi }_{1}}{{\alpha}_{m,n}}+\sum\limits_{n\in \mathcal{N}}{{{\Xi }_{2}}{{\alpha}_{m,n}}}+W{{\xi }_{0}}\right)}_{p_{m}(\alpha )}
-\\\underbrace{W{{\log }_{2}}\left(\sum\limits_{n\in \mathcal{N}}{{{\Xi }_{2}}{{\alpha}_{m,n}}}+W{{\xi }_{0}}\right)}_{q_{m}(\alpha )},
\end{split}
\end{equation}
where ${{\Xi }_{1}}={{P}_m}|{{h}_{m,B}}+{{\textbf h}_{R,B}}{\bm{\Phi }_{r}}{{\textbf h}_{m,R}}{{|}^{2}}$, ${{\Xi }_{2}}=|{{h}_{n,B}}{{|}^{2}}$. The former part is concave while the latter is convex. In this context, we contemplate employing the Successive Convex Approximation (SCA) technique \cite{SCA} at each iteration to approximate the original non-convex problem at a given local point. In the $k$th interation, we define ${\bm{\alpha}^{(k)}}=\{{\alpha}_{m,n}^{(k)}, \forall m\}$ as the expansion of the specific point.

Since the first half becomes concave after the transformation, the DC variant of the problem remains unsolvable. To address this, we rewrite it as a convex function via a first-order Taylor expansion at a specific point. This enables us to obtain an upper bound on the objective function and facilitates the derivation of the optimal solution. Subsequently, the first-order Taylor expansion of ${{q}_{m}}(\bm{\alpha})$ at the given point ${{\bm{\tilde{\alpha}}\,}}$ in the $k$th iteration is obtained as follow
\begin{equation}
\begin{split}
&q_{_{m}}(\bm{\alpha})\!=\!{{q}_{m}}({\bm{\alpha}^{(k)}})\!+\!\sum\limits_{n\in \mathcal{N}}{{{\left. \frac{\partial {{q}_{m}}(\bm{\alpha})}{\partial {{\alpha}_{m}}} \right|}_{\bm{\alpha}={{\alpha}^{(k)}}}}}\left({{{\tilde{\alpha}}\,}_{m,n}}-{\tilde{\alpha}}\,_{_{m,n}}^{(k)}\right).\\
%& \frac{\partial {{q}_{m}}(\alpha )}{\partial {{\alpha }_{m}}}=\frac{2}{\left(\sum\limits_{n\in \mathcal{N}}{{{\Xi }_{2}}{{\alpha }_{m,n}}+W{{\xi }_{0}}} \right)\ln 2}\left[ \sum\limits_{n\in \mathcal{N}}{{\textbf h}_{n,B}^{H}{{\textbf h}_{n,B}}{{\alpha }_{m,n}^{(k)}}} \right], \\ 
%&  {{A}_{m}}\left( {{\alpha }_{m,n}^{(k)}} \right)=\sum\limits_{n\in \mathcal{N}}{{{\Xi }_{2}}{{\alpha }_{m,n}}+W{{\xi }_{0}}}. \\
\end{split} 
\end{equation}

Substituting into the original equation we can obtain that
\begin{equation}
\begin{aligned}
{\mathop{\tilde{R}_{B}^{m}}}(\bm{\alpha}^{(k)})=&{{p}_{m}}(\bm{\alpha})-{{q}_{m}}({\bm{\alpha}^{(k)}})
-\\&\sum\limits_{n\in \mathcal{N}}{{{\left. \frac{\partial {{q}_{m}}(\bm{\alpha})}{\partial {{\alpha}_{m}}} \right|}_{\bm{\alpha}={{\alpha}^{(k)}}}}}\left({{{\tilde{\alpha}}\,}_{m,n}}-{\tilde{\alpha}}\,_{_{m,n}}^{(k)}\right).
\end{aligned}
\end{equation}

%\begin{figure*}[t]
%\begin{equation} \label{gamma}
%\begin{aligned}
%{{\gamma }_{const}}\left( {\sum\nolimits_{m=1}^{M}{{{\alpha }_{m,n}}}{{P}_{m}}\left( {{\Gamma }_{2}}\left( |\sqrt{{{\kappa }_{R,n}}{{\kappa }_{m,R}}}\cdot {\textbf h}_{R,n}^{\rm LoS} \mathbf{\Phi}_t {\textbf h}_{m,R}^{\rm LoS}{{|}^{2}}+{{\Gamma }_{1}} %\right) \right)+W{{\xi }_{0}}}\right) \times {{{\tilde{P}}\,}_{t}}\le {{{\tilde{P}}\,}_{m}}
%\end{aligned}
%\end{equation}  
%\end{figure*}

Thus the problem (\ref{tau}) can be reformulated as follows
\begin{subequations} \label{alpha}
\begin{align}
 &\underset{\bm{\alpha}}{\mathop{\min }}\;\;\;\;\;\sum\limits_{m\in \mathcal{M}}{\tau _{m}^{(k)}} \nonumber \\ 
 &{{s}}{{.t}}{\rm{.}}\;\;\;\;\;\;\;(\ref{reformulate}a), (\ref{equivalent_problem}a), (\ref{binary_original_problem}c). \nonumber
\end{align}
\end{subequations}

At this point, the objective function and the constraint are convex, so we can solve them by using the standard convex optimization solver such as CVX~\cite{CVX}.

\subsection{RICS's Reflection and Refraction Optimization}
Given $\bm{\rho}$ and $\bm{\alpha}$, the original problem can be restructured to minimize the sum of the delay. As the delay function incorporates a maximum operator, where the first half is constant, the entire function is solely impacted by the changes in $R_{B}^{m}$. Therefore, our ultimate goal is to determine a distribution of ${\mathbf{\Phi}}_x$ that maximizes the sum of $R_{B}^{m}$.

Since both the objective function and the constraint (\ref{binary_original_problem}d)- (\ref{binary_original_problem}e) are non-convex, the origin problem (\ref{equivalent_problem}) is still non-convex with respect to $\bm{\Phi}_x$. To solve this problem, we first have to ensure that the problem under consideration is convex. Therefore, in this scenario, we will utilize the Semidefinite Relaxation (SDR)\cite{SDR} technique to relax the rank constraint and transform the original problem into a convex optimization problem. 

Given that the optimization variables $\bm{\Phi}_{x},x \in \{t,r\}$ only appear in the SINR $\gamma_{B}^{m}$ and the constraint (\ref{binary_original_problem}d)-(\ref{binary_original_problem}e). Considering the molecular component of the safety coefficient ${S}_{m}$ remains constant due to the fixation of $\bm{\rho}$, so we replanned the problem as follow
\begin{subequations} \label{phi_reformulate}
\begin{align}
& \underset{{{\bm{\Phi} }_{t}},{{\bm{\Phi} }_{r}}}{\mathop{\max }}\,\text{    }\sum\limits_{m\in \mathcal{M}}{{{\log }_{2}}\left(1+\frac{{{P}_{m}}|{h}_{m,B} \!+\! {\textbf h}_{R,B} \mathbf{\Phi}_t {\textbf h}_{m,R}\big|^2}{\sum\nolimits_{n=1}^{N}{{{\alpha }_{m,n}}{{P}_{t}}|{{h}_{n,B}}{{|}^{2}}+W{{\xi }_{0}}}}\right)} \nonumber \\
&{{s}}{{.t}}{\rm{.}}%\;\;\;\;\;\;\;\;\;\;\;\;\theta _{l}^{r},\theta _{l}^{t}\in [0,2\pi ),1\le l\le L, \\ 
%&\;\;\;\;\;\;\;\;\;\;\;\;\;\;\;\;\;\;\;\;\beta _{l}^{r}+\Psi_{l}\beta _{l}^{t}=1,1\le l\le L, \\
\;\;\;\;\;\;(\ref{binary_original_problem}d), (\ref{binary_original_problem}e). \nonumber
\end{align}
\end{subequations}

To better handle the terms in the SINR expression, $|{{h}_{m,B}}+{\textbf{h}_{R,B}}{\bm{\Phi}_{r}}{\textbf{h}_{m,R}}{{|}^{2}}$, we introduce the vector ${{\bm{\Theta}}_{x}}\in {{\mathbb{C}}^{L\times 1}}$ as the main diagonal element of $\bm{\Phi }_{x}$, that is ${{\bm{\Theta}}_{x}}={{[s_{1}^{x},\ldots,s_{L}^{x}]}^{T}}, \forall x\in \{t,r\}$. Then, the auxiliary variables ${{\textbf{h}}_{B}}\in {{\mathbb{C}}^{L\times 1}}$ are introduced and there is ${{\textbf{h}}_{B}}={\textbf{h}_{R,B}}\circ {\textbf{h}_{m,R}}$.
So the original problem can be rewritten as
\hspace{-2mm}
\begin{subequations} \label{phi_SDR}
\begin{align}
& \underset{{\bm{\Theta}}_{t},{\bm{\Theta}}_{r}}{\mathop{\max }}\,\text{    }\sum\limits_{m\in \mathcal{M}}{{{\log }_{2}}\left(1+\frac{{{P}_{m}}|{{h}_{m,B}}+\bm{\Theta}_{r}^{H}{{{\textbf h}}_{B}}{{|}^{2}}}{\sum\nolimits_{n=1}^{N}{{{\alpha }_{m,n}}{{P}_{t}}|{{h}_{n,B}}{{|}^{2}}+W{{\xi }_{0}}}}\right)} \nonumber \\ 
& {{s}}{{.t}}{\rm{.}}\text{      }{\sum\limits_{m\in \mathcal{M}}({{{\alpha }_{m,n}}{{P}_{m}}|{{h}_{m,n}}+\bm{\Theta}_{t}^{H}{{{\textbf h}}_{n}}{{|}^{2}}+W{{\xi }_{0}}}})\leq \frac{{{P}_{t}}|{\bm{h}_{n}}{{|}^{2}}}{\tilde{\gamma }_{c}},  \\ 
&\;\;\;\;\;\;\;\;\;\;{{\left. \left| \left[ \bm{\Theta} _{t}^{l} \right] \right. \right|}^{2}}+{{\left. \left| \left[ \bm{\Theta} _{r}^{l} \right] \right. \right|}^{2}}=1,\forall l\in \{1,2,\ldots,L\}.
\end{align}
\end{subequations}

Next, we introduce the auxiliary variables $\bm{R}$ and $\bm{v}$
\begin{equation*}
{\bm{R}_{B}}={{P}_{m}}\left[ 
\begin{aligned}
  & {\textbf{h}_{B}}\textbf{h}_{B}^{H}\;\;\;\;\;{\textbf{h}_{\text{B}}}h_{m,B}^{H} \\ 
 & {\textbf{h}_{B}^{H}}{h_{m,B}}\text{     }h_{m,B}^{H}{h_{m,B}} \\ 
\end{aligned} \right],{{\overset{\_}{\mathop{\bm{v}}}\,}_{x}}=\left[ 
\begin{aligned}
  & {\bm{\Theta}_{x}} \\ 
 & 1 \\ 
\end{aligned} \right],\forall x\in \{t,r\}.
\end{equation*}

Thus, we have the relation as follows
\begin{equation}
{{P}_{m}}|{h_{m,B}}+\bm{\Theta} _{x}^{H}{\textbf{h}_{B}}{{|}^{2}}=\overline{\bm{v}}_{x}^{H}{\bm{R}_{B}}{\overline{\bm{v}}_{x}}.
\end{equation}

Due to ${{{{\overline{\bm{v}}_{x}^{H}}}}}{\bm{R}_{m}}{{\overline{\bm{v}}_{x}}}=\text{Tr}({\bm{R}_{m}}{{\overline{\bm{v}}_{x}}}{{{{\overline{\bm{v}}_{x}^{H}}}}})$, we define ${\bm{V}_{x}}={{\overline{\bm{v}}_{x}}}{{{{\overline{\bm{v}}_{x}^{H}}}}}$, which satisfy ${\bm{V}_{x}}\ge 0$ and rank$({\bm{V}_{x}})=1$, then the original problem is stated as follows
\begin{subequations} \label{problem36}
\begin{align}
& \underset{{\bm{V}_{r}},{\bm{V}_{t}}}{\mathop{\max }}\,\text{    }\sum\limits_{m\in \mathcal{M}}{{{\log }_{2}}\left(1+\frac{Tr\left({\bm{V}_{r}}{\bm{R}_{B}}\right)}{\sum\nolimits_{n=1}^{N}{{{\alpha }_{m,n}}{{P}_{t}}|{{h}_{n,B}}{{|}^{2}}+W{{\xi }_{0}}}}\right)}, \nonumber \\ 
 & {{s}}{{.t}}{\rm{.}}\;\;\;\text{      }{\sum\nolimits_{m=1}^{M}({{{\alpha }_{m,n}}Tr({\bm{V}_{t}}{\bm{R}_{n}})+W{{\xi }_{0}}}})\leq \frac{{{P}_{t}}|{{h}_{n}}{{|}^{2}}}{\tilde{\gamma }_{c}}, \\ 
 &\;\;\;\;\;\;\; \;\;\;{{\left[ {\bm{V}_{r}} \right]}_{l,l}}+{{\left[ {\bm{V}_{t}} \right]}_{l,l}}=1,{{\left[ {\bm{V}_{x}} \right]}_{l,l}}\ge 0,\forall l\in \{1,2,\ldots,L\},  \\ 
 & \;\;\;\;\;\;\;\; \;\;{{\left[ {\bm{V}_{r}} \right]}_{L+1,L+1}}=1,{\bm{V}_{r}}\succcurlyeq 0,{\bm{V}_{t}}\succcurlyeq 0,\\
&\;\;\;\;\;\;\;\;\;\;\; rank({\bm{V}_{r}})=rank({\bm{V}_{t}})=1, \\
&\;\;\;\;\;\;\;\;\;\;\;(\ref{binary_original_problem}d), (\ref{binary_original_problem}e). \nonumber
\end{align}
\end{subequations}

%Where, ${{\left[ \cdot  \right]}_{l,l}}$ denotes the lth row and lth column elements of the matrix. In the ablove question,the constraint (\ref{problem36}c) is still not convex, and we first relax the rank-first constraint,then the problem becomes convex. Since this is a convex semidedefinite programming (SDP), we use the standard convex optimization solver CVX to solve it, and then use the Gaussian randomization process to recover $\bm{\Theta}$ and obtain the corresponding $\bm{\Theta}^{opt}_{x}$. Then we have
%\begin{equation}
%\bm{\Phi} _{x}^{opt}=diag\left( \bm{\Theta} _{x}^{opt} \right)
%\end{equation}

It is clear that the constraints (\ref{binary_original_problem}d)-(\ref{binary_original_problem}e) are non-convex, so we deal with this by introducing an auxiliary variable ${\mathop{{\tilde{\bm{v}}_{x}}}}\,={{\left[\sqrt{\tilde{\beta} _{1}^{x}}{{e}^{j\tilde{\theta} _{1}^{x}}},\ldots,\sqrt{\tilde{\beta} _{L}^{x}}{{e}^{j\tilde{\theta} _{L}^{x}}}\right]}^{T}}$. In order to avoid falling into local optimal and obtain more accurate solutions, we establish a penalty convex relaxation problem to relax the coupled phases, where we construct an Augmented Lagrangian (AL) problem by using ${{\tilde{\mathop{\bm{v}}}\,}_{x}}={\bm{v}_{x}}$ as a penalty term
\begin{subequations} \label{coupled}
\begin{align}
  & \underset{{\bm{V}_{t}},{\bm{V}_{r}}}{\mathop{\max }}\,\;\;\;\;\bm{\vartheta}  \nonumber\\ 
 & s.t.\;\;\;\;{{\tilde{\mathop{\bm{v}}}\,}_{r}}(l)=\pm j{{\tilde{\mathop{\bm{v}}}\,}_{t}}(l),{{\sum\limits_{x\in \{t,r\}}{\left| {{\tilde{\mathop{\bm{v}}}\,}_{x}}(l) \right|}}^{2}}=1, \\
 & \;\;\;\;\;\;\;\;\;(\ref{problem36}b)-(\ref{problem36}d), \nonumber
\end{align}
\end{subequations}
where $\eta >0$ holds, $\bm{\vartheta} =\overset{-}{\mathop{R}}\,_{B}^{m}-\frac{1}{2\eta }\left( {{\sum\limits_{x\in \{t,r\}}{\left\| {{\tilde{\mathop{\bm{v}}}\,}_{x}}\tilde{\mathop{\bm{v}}}\,_{x}^{H}-{\bm{V}_{x}} \right\|}}^{2}} \right)$.

However, due to the presence of $rank({\bm{V}_{x}})=1$, we can equivalently transform them into $Tr\left( {\bm{V}_{x}} \right)={{\left\| {\bm{V}_{x}} \right\|}_{2}}$. Then, we use a penalty-based DC method to remove the rank-one constraint as follows
\begin{equation}
{\tilde{\bm{\vartheta }}^{p}}=\bm{\vartheta} -\varpi \left( \sum\limits_{x\in \{t,r\}}{\Re \left( Tr\left( \bm{V}_{x}^{H}\left( \bm{I}-{\bm{v}_{x,1}}\bm{v}_{x,1}^{H} \right) \right) \right)} \right),
\end{equation}
where $\varpi >0$ stands for the rank-one penalty coefficient. When $\varpi >$ approaches $+\infty$, we can obtain the rank one ${\bm{V}_{x}}$.

For the optimization of coupled reflection and refraction coefficient matrix, we refered to the method of \cite{CoupledPhase}. For the fixed rank-one matrix ${\bm{V}_{x}}$, we have that ${{\left\| {{\tilde{\mathop{\bm{v}}}\,}_{x}}\tilde{\mathop{\bm{v}}}\,_{x}^{H}-{\bm{V}_{x}} \right\|}^{2}}$ can be reduced to ${{\left\| {{\tilde{\mathop{\bm{v}}}\,}_{x}}\tilde{\mathop{\bm{v}}}\,_{x}^{H}-{\bm{v}_{x}}\bm{v}_{x}^{H} \right\|}^{2}}$. The original optimization problem can be rewritten as
\begin{subequations} \label{DC}
\begin{align}
  & \underset{{{\tilde{\mathop{\bm{v}}}\,}_{x}}}{\mathop{\min }}\,{{\sum\limits_{x\in \{t,r\}}{\left\| {{\tilde{\mathop{\bm{v}}}\,}_{x}}-{\bm{v}_{x}} \right\|}}^{2}}\nonumber \\ 
 & s.t. \;\;\;\;\;(\ref{coupled}a).\nonumber
\end{align}
\end{subequations}

Then we expand the problem (\ref{DC}) and obtain that
\begin{equation}
\begin{split}
   {{\left\| {{\tilde{\mathop{\bm{v}}}\,}_{x}}-{\bm{v}_{x}} \right\|}^{2}}&={{\left( {{\tilde{\mathop{\bm{v}}}\,}_{x}}-{\bm{v}_{x}} \right)}^{H}}\left( {{\tilde{\mathop{\bm{v}}}\,}_{x}}-{\bm{v}_{x}} \right) \\ 
 & =\left( {{\tilde{\mathop{\bm{v}}}\,}_{x}}^{H}{{\tilde{\mathop{\bm{v}}}\,}_{x}} \right)-2\Re \left( \bm{v}_{x}^{H}{{\tilde{\mathop{\bm{v}}}\,}_{x}} \right)+\bm{v}_{x}^{H}{\bm{v}_{x}}.
\end{split}
\end{equation}

Since the real part of the inner product of ${\tilde{\mathop{\bm{v}}}\,}_{x}$ and $\bm{v}_{x}$ is proportional to the objective function, maximizing the real part of the inner product minimizes the objective function. Thus the opimization problem (\ref{DC}) can be rewritten as
\begin{align}\label{reformulateDC}
&\underset{{\tilde{\bm{v}}_{x}}\left( l \right)}{\mathop{\max }}\,\;\;\;\Re \left( \bm{v}_{t}^{H}\left( l \right){\tilde{\bm{v}}_{t}}\left( l \right) \right)+\Re \left( \bm{v}_{r}^{H}\left( l \right){\tilde{\bm{v}}_{r}}\left( l \right) \right)\\ 
& s.t.\;\;\;\;\;\left( \ref{coupled}a \right).\nonumber
\end{align}

Let $\bm{u}_{x}^{H}\!=\!\bm{v}_{x}^{H}diag\left( \left[ \sqrt{\tilde{\beta}_{1}^{x}},\ldots,\sqrt{\tilde{\beta} _{L}^{x}} \right] \right)$ when  $\tilde{\beta} _{l}^{x}$ is given, then the problem (\ref{reformulateDC}) is equivalent to $\Re \left( \left[ \bm{v}_{t}^{H}(l)\pm j\bm{v}_{t}^{H}(l) \right]{{e}^{j\tilde{\theta} _{l}^{t}}} \right)$. Until when $\angle \left( \left[ \bm{u}_{t}^{H}(l)\pm j\bm{u}_{t}^{H}(l){{e}^{j\tilde{\theta} _{l}^{t}}} \right] \right)=0$ holds, it can reach the maximum. Meanwhile, for the given $\tilde{\theta} _{l}^{x}$, the problem (\ref{reformulateDC}) is equivalent to $\Re \left( \bm{\varphi} _{t}^{H}(l)\sqrt{\tilde{\beta} _{l}^{t}}+\bm{\varphi} _{r}^{H}(l)\sqrt{\tilde{\beta} _{l}^{r}} \right)$, where $\bm{\varphi} _{x}^{H}\!=\!\bm{v}_{x}^{H}diag\left( \left[ {{e}^{j\tilde{\theta} _{1}^{x}}},\ldots,{{e}^{j\tilde{\theta} _{L}^{x}}} \right] \right)$ holds. Let ${{a}_{n}}\!=\!\left| \bm{\varphi} _{t}^{H}(l) \right|\cos (\angle \bm{\varphi} _{t}^{H}(l)),$ $b_n=\left| \bm{\varphi} _{r}^{H}(l) \right|\cos (\angle \bm{\varphi} _{r}^{H}(l))$, we can finally reformulate the original optimize problem as ${{a}_{n}}\sqrt{\tilde{\beta} _{l}^{t}}+{{b}_{n}}\sqrt{\tilde{\beta} _{l}^{r}}.$
Note that problem (\ref{DC}) can be solved by separately solving L-independent subproblems to obtain optimal coupled amplitude and phase shift solutions.
During the iterative solving process of the problem (\ref{coupled}), the value of $\bm{v}$ is updated using (\ref{theta}) and (\ref{beta}). Subsequently, the updated values are substituted into the problem (\ref{DC}) to obtain feasible solutions $\bm{v}_{t}$ and $\bm{v}_r$. By comparing the objective function values, the larger one is chosen as the solution for the current iteration.
\begin{equation} \label{theta}
\tilde{\theta} _{l}^{t}=\left\{ \begin{aligned}
  & -\angle \left( \bm{u}_{t}^{H}\left( l \right)+j\bm{u}_{t}^{H}\left( l \right) \right),\left( \tilde{\theta} _{l}^{r}=\tilde{\theta} _{l}^{t}+\frac{\pi }{2} \right) \\ 
 & -\angle \left( \bm{u}_{t}^{H}\left( l \right)+j\bm{u}_{t}^{H}\left( l \right) \right),\left( \tilde{\theta} _{l}^{r}=\tilde{\theta} _{l}^{t}+\frac{3\pi }{2} \right) \\ 
\end{aligned}. \right. 
\end{equation}
\begin{equation} \label{beta}
\left\{ \begin{aligned}
  & \sqrt{\tilde{\beta} _{l}^{t}}=\frac{{{a}_{l}}}{\sqrt{a_{l}^{2}+b_{l}^{2}}},\sqrt{\tilde{\beta} _{l}^{r}}=\frac{{{a}_{l}}}{\sqrt{a_{l}^{2}+b_{l}^{2}}},if\text{  } {{p}_{n}},{{q}_{n}}\ge 0. \\ 
 & \sqrt{\tilde{\beta} _{l}^{t}}=1,\sqrt{\tilde{\beta} _{l}^{r}}=0,\text{ }if\text{ }{{p}_{n}}\ge 0,\text{ }{{q}_{n}}<0. \\ 
 & \sqrt{\tilde{\beta} _{l}^{t}}=0,\sqrt{\tilde{\beta} _{l}^{r}}=1,\text{ }if\text{ }{{p}_{n}}<0,\text{ }{{q}_{n}}\ge 0. \\ 
 & \sqrt{\tilde{\beta} _{l}^{t}}=0,\sqrt{\tilde{\beta} _{l}^{r}}=0,\text{ }otherwise. \\ 
\end{aligned} \right.
\end{equation}

By alternately optimizing the above three variables, we propose an Alternative Iterative Optimization Algorithm ({AIOA}) framework, shown in \textbf{Algorithm~\ref{algorithm2}}.
\begin{algorithm}[t]
\caption{The proposed {AIOA} framework for solving problem (\ref{equivalent_problem})}\label{algorithm2}
\begin{algorithmic}
\STATE 
\STATE Initialize   ${\mathbf{\bm{\alpha}^{(0)}}}$, ${\mathbf{\bm{\Phi}^{(0)}}}$;
\STATE $k=0$;
\WHILE{$\sum\limits_{m\in \mathcal{M}}{\left( S_{m}^{(k+1)}-S_{m}^{(k)} \right)/}S_{m}^{(k+1)}\ge \delta$ }
    \STATE Update ${\mathbf{\bm{\rho}^{(k+1)}}}$ by using \textbf{Algorithm}~\ref{a1};
    \STATE Solve the problem (\ref{alpha}) by using CVX for given \{${\mathbf{\bm{\rho}^{(k+1)}}}$, ${\mathbf{\bm{\alpha}^{(k)}}}$\} and obtain the solution \{${\mathbf{\bm{\rho}^{(k+1)}}}$,${\mathbf{\bm{\alpha}^{(k+1)}}}$, ${\mathbf{\bm{\Phi}^{(k)}}}$\};
    \STATE For the given \{${\mathbf{\bm{\rho}^{(k+1)}}}$, ${\mathbf{\bm{\alpha}^{(k+1)}}}$, ${\mathbf{\bm{\Phi}^{(k)}}}$\}, using Equation (\ref{theta})(\ref{beta}) to solve problem (\ref{coupled}) to obtain \{${\mathbf{\bm{\Phi}^{(k+1)}}}$ \};
	\STATE {$k=k+1$;}	 
\ENDWHILE
\STATE \textbf{return} {${\mathbf{\bm{\rho}^{(*)}}}$, ${\mathbf{\bm{\alpha}^{(*)}}}$ and ${\mathbf{\bm{\Phi}^{(*)}}}$.}
\end{algorithmic}
\end{algorithm}

\begin{algorithm}[t]
\caption{Gradient descent-based approach for solving problem (\ref{objective_func})}\label{algorithm3}
\begin{algorithmic}
\STATE 
\STATE Initialize   $f(\bm{\Psi})$;
\STATE $k=0$;
\WHILE {$\left\| \nabla f\left( {\bm{\Psi }^{(t+1)}} \right) \right\|\ge \epsilon $ }
		\STATE Caculate gradient $\nabla f\left( {\bm{\Psi }^{(t)}} \right)$;
		\STATE Update ${\bm{\Psi }^{(t+1)}}={\bm{\Psi }^{(t)}}-{{a}^{(t)}}\nabla f\left( {\bm{\Psi }^{(t)}} \right)$;\\
		\STATE Update the learning rate $a$ using the linear search method;
		\STATE {$t=t+1$;}	  
\ENDWHILE
\end{algorithmic}
\end{algorithm}

\subsection{Optimal Amplitude Adjustment Factor Exploration}
For the joint optimization of offloading ratio $\bm{\rho}$, the spectrum sharing strategy $\bm{\alpha}$ and the reflection and refraction matrices $\bm{\Phi}_x$, we need to determine a suitable adjustment factor of the RICS to maximize the safety coefficient of the CVs. Notably, an appropriate adjustment factor $\bm{\Psi}$ should be achieved to effectively mitigate the interference between the CVs and the $R_x$ of the V2V pair, thereby enhancing its data rate. Therefore, our goal is 
\begin{equation}
\underset{\bm{\Psi }}{\mathop{\min }}\,\sum\limits_{m\in \mathcal{M}}{{{\alpha }_{m,n}}}{{P}_{m}}{{\left| {{h}_{m,n}}+\textbf{h}_{R,n}^{H}{\bm{\Phi }_{t}}{\textbf{h}_{m,R}} \right|}^{2}},
\end{equation}
where $\left| \cdot  \right|$ indicates that each item from 1 to ${M}$ is greater than or equal to 0. 

Unlike the method of solving the reflection and refraction coefficient matrices mentioned above, we have now constructed a Quadratic Programming (QP) problem in which the optimization variables $\bm{\Psi}$ are convex. Therefore, we consider using gradient descent to find the optimal distribution of  $\bm{\Psi}$. This method aims to minimize a convex objective function through iterative steps in the direction of the negative gradient.
For the fixed $\bm{\alpha}$ and transmission power ${P}_m$, we need to investigate the values of the vectors $\bm{\Psi} ={{[{{\Psi }_{1}},\ldots,{{\Psi }_{l}}]}^{T}}$, so our optimization problem can be equivalently expressed as
\begin{equation}
\underset{\bm{\Psi }}{\mathop{\min }}\sum\limits_{m\in \mathcal{M}}{\left( {{h}_{m,n}}+\sum\limits_{i=1}^{L}{\textbf{h}_{R,n}^{H}{{\Psi }_{i}}\sqrt{{{\beta }_{i}}}{{e}^{j{{\theta }_{i}}}}}{\textbf{h}_{m,R}} \right)}.
\end{equation}

%That is, for each CV, to ensure ${{h}_{m,n}}=-\sum\limits_{i=1}^{L}{\textbf{h}_{R,n}^{H}{{\Psi }_{i}}\sqrt{{{\beta }_{i}}}{{e}^{j{{\theta }_{i}}}}}{\textbf{h}_{m,R}}$, that we can obtain the nonlinear systems of equations
%\begin{equation}
%\left\{ \begin{aligned}
%  & {\Re}({{h}_{m,n}})=-\sum\limits_{i=1}^{L}{{{\Psi }_{i}}}{\Re}(\textbf{h}_{R,n}^{H}\sqrt{{{\beta }_{i}}}\cos ({{\theta }_{i}}){\textbf{h}_{m,R}}) \\ 
% & {\Im}({{h}_{m,n}})=-\sum\limits_{i=1}^{L}{{{\Psi }_{i}}}{\Im}(\textbf{h}_{R,n}^{H}\sqrt{{{\beta }_{i}}}j\sin ({{\theta }_{i}}){\textbf{h}_{m,R}}) \\ 
%\end{aligned} \right.
%\end{equation}

Since only one unknown is left, we apply the least squares method, where the objective is to minimize the sum of the residual squares to obtain the optimal value of $\bm{\Psi}$. So we rewrite the objective function as
\begin{equation} \label{objective_func}
\begin{split}
\underset{\bm{\Psi}}{\mathop{\min }}\, \sum\limits_{m\in \mathcal{M}}{{\left( \Re({{h}_{m,n}})+\sum\limits_{i=1}^{L}{{{\Psi }_{i}}}\Re\left( \textbf{h}_{R,n}^{H}\sqrt{{{\beta }_{i}}}\cos ({{\theta }_{i}}){\textbf{h}_{m,R}} \right) \right)}^{2}}+\\
{{\left( \Im({{h}_{m,n}})+\sum\limits_{i=1}^{L}{{{\Psi }_{i}}}\Im\left( \textbf{h}_{R,n}^{H}\sqrt{{{\beta }_{i}}}j\sin ({{\theta }_{i}}){\textbf{h}_{m,R}} \right) \right)}^{2}}.
\end{split}
\end{equation}

\begin{figure*}[!h] 
	\centering
    \includegraphics[width=1.5\columnwidth]{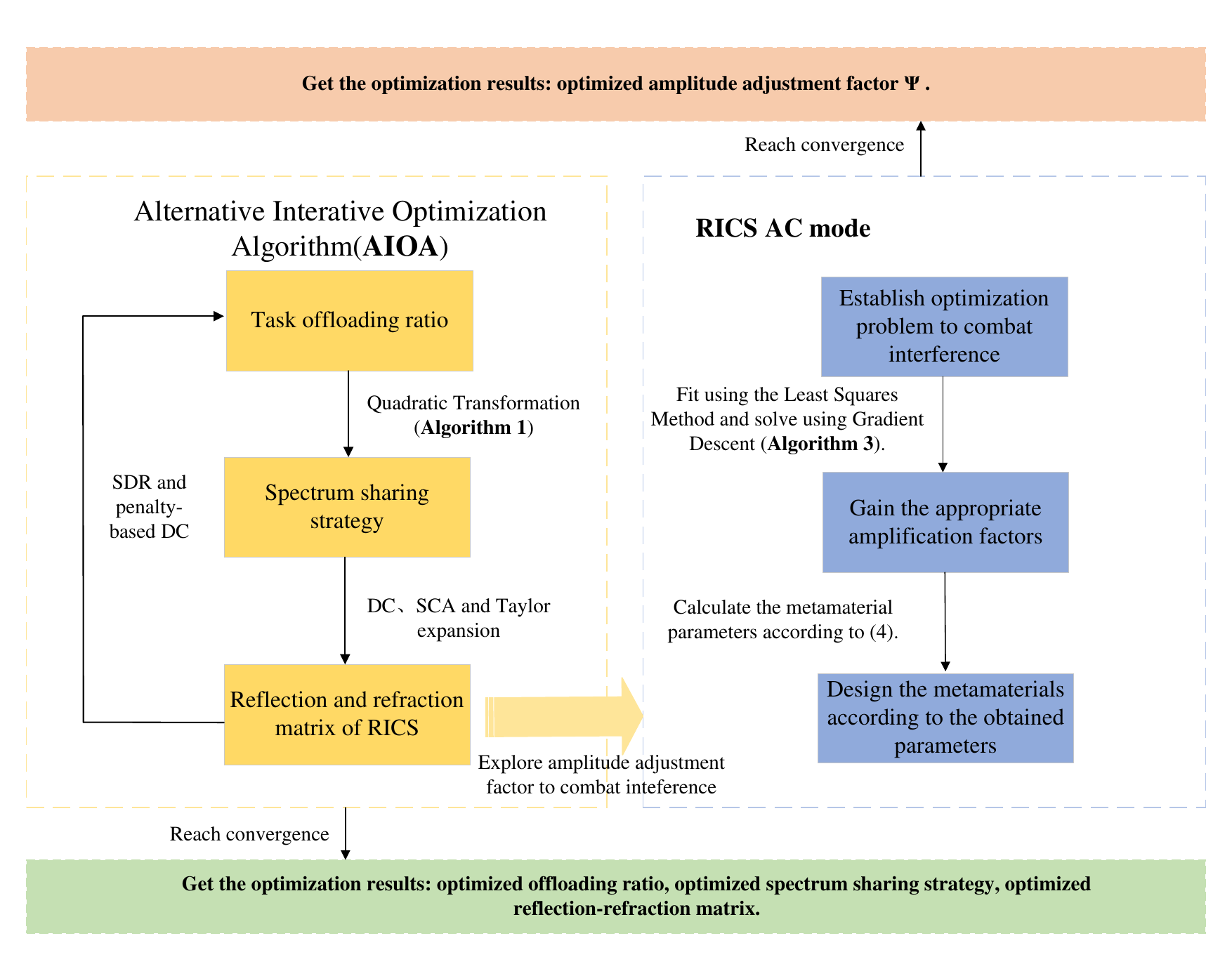}
    \caption{The framework of the proposed algorithm.}
    \label{framework}
\end{figure*}

The above is a nonlinear least squares problem of $\bm{\Psi}$. To solve this problem, we employ the gradient descent method, as shown in \textbf{Algorithm~\ref{algorithm3}}, where the partial derivative is taken for ${\Psi}_i$ and the $i$th component of its gradient vector is denoted as $\nabla f\left( {{\Psi}_{i}} \right)$.

The core idea of the algorithm is to start from an initial point and take a small step in the opposite direction of the function's gradient at each step, gradually approaching the minimum value of the function through iteration. The choice of the initial point is crucial as it affects both the convergence speed and stability of the algorithm. Here, we choose the initial point as a row vector with all elements of $\bm{\Psi}$ set to 1. This initial value provides us with a relatively neutral starting point, assisting the algorithm in approaching the optimal solution more quickly. Additionally, it mitigates the risk of premature convergence to suboptimal solutions due to varying adjustment weight coefficients of different RICS elements.
Moreover, the learning rate $a$ is also an important parameter, which determines the step size of each update. Considering that its selection introduces a trade-off between the convergence speed and stability of the algorithm, to balance accuracy and speed, we set $a$ to 0.01. This setting helps to ensure the quality of the solution while avoiding stability issues caused by excessive adjustments in the algorithm.
%In Section \ref{sec:algorithm-three-b}, we mentioned that the parameter design of MS in the analog computation mode is influenced by the transfer function, which is related to $\bm{\Psi}$. Since different RICS elements have different amplification factors. In this section, with the help of Algorithm \ref{algorithm3}, we obtain the values of the amplification factor $\bm{\Psi}$, allowing us to adjust the various parameters of the analog computation mode's composite materials adaptively based on the amplification coefficients. 
 For better understanding, we present the proposed algorithm framework in the form of a flowchart, as shown in Fig.~\ref{framework}.\vspace{-1.5mm}
\subsection{Complexity Analysis}
%For solving the subproblems with the above three optimization variables, we propose an alternating optimization framework to approximate the optimal solution of the original problem. This method works by updating only one variable at one time in each iteration while keeping the other varibles fixed. Through iteratively updating all varibles in a loop until a convergence condition is reached.
In this part, we analyze the convergence of the proposed algorithmic framework {AIOA}. Firstly, we define the optimal solution of the algorithm at the $k$th iteration $\Gamma \left( {\bm{\rho }^{\left( k \right)}},{\bm{\alpha }^{\left( k \right)}},{\bm{\Phi }^{\left( k \right)}} \right)$. In step 4 of the \textbf{Algorithm~\ref{algorithm2}}, we optimize the problem (\ref{Quad}) with the given $\left\{ {\bm{\rho }^{(k+1)}},{\bm{\alpha }^{(k)}} \right\}$. Thus, we have the relation
\begin{equation}
\begin{split}
\Gamma \left( {\bm{\rho }^{(k+1)}},{\bm{\alpha }^{(k)}},{\bm{\Phi }^{(k)}} \right)\overset{\left( a \right)}{\mathop{=}}\,\Gamma _{\bm{\alpha} }^{ub}\left( {\bm{\rho }^{(k+1)}},{\bm{\alpha }^{(k)}},{\bm{\Phi }^{(k)}} \right)\\
\overset{(b)}{\mathop{\ge }}\,\Gamma _{\bm{\alpha} }^{ub}\left( {\bm{\rho }^{(k+1)}},{\bm{\alpha }^{(k+1)}},{\bm{\Phi }^{(k)}} \right),
\end{split}
\end{equation}
where (a) indicates that we employed the first-order Taylor expansion at the given point to approximate the original optimal solution. (b) denotes the utilization of the log-sum-exp to  convert the original non-convex problem to convex, which can be regarded as an upper bound on the original optimal solution.

Therefore, the absolute value of the optimal solution to the problem (\ref{alpha}) that we obtain can provide a lower bound for the original problem (\ref{reformulate}). We have an inequality relation 
\begin{equation}
\left| \Gamma _{\bm{\alpha} }^{ub}\left( {\bm{\rho }^{(k+1)}},{\bm{\alpha }^{(k+1)}},{\bm{\Phi }^{(k)}} \right) \right|\le \Gamma \left( {\bm{\rho }^{(k+1)}},{\bm{\alpha }^{(k+1)}},{\bm{\Phi }^{(k)}} \right).
\end{equation}

From above, we can conclude that in solving the subproblem of optimizing the variable $\bm{\alpha}$, its solution is non-increasing during the iteration process. As indicated in step 6 of the \textbf{Algorithm~\ref{algorithm2}}, when we solve the reflection and refraction coefficient matrix of RICS, the solution obtained is optimal, so we can obtain that %$\Gamma \left( {\bm{\rho }^{(k+1)}},{\bm{\alpha }^{(k+1)}},{\bm{\Phi }^{(k)}} \right)\le \Gamma \left( {\bm{\rho }^{(k+1)}},{\bm{\alpha }^{(k+1)}},{\bm{\Phi }^{(k+1)}} \right)$.
%Based on the above analysis, we can obtain that
\begin{equation} \label{converge}
\Gamma \left( {\bm{\rho }^{(k)}},{\bm{\alpha }^{(k)}},{\bm{\Phi }^{(k)}} \right)\le \Gamma \left( {\bm{\rho }^{(k+1)}},{\bm{\alpha }^{(k+1)}},{\bm{\Phi }^{(k+1)}} \right),
\end{equation}
where we can guarantee that the algorithm converges to a fixed value. Next, we analyze the complexity of the {AIOA}.

%The alternating optimization algorithm proposed in this paper contains stepwise optimization of three variables. 
First of all, the problem complexity of optimizing the offloading ratio of the CVs using the Quadratic Transform algorithm is $\mathcal{O}\left( {{\textit{K}}{M}} \right)$, where $K$ is the number of \textbf{Algorithm~\ref{a1}}. Then the complexity of optimizing the spectrum sharing strategy is $\mathcal{O}\left( {{\textit{M}}{N}} \right)$. Moreover, the standard convex SDP optimization problem is used for solving the reflection and refraction coefficient matrices of the RICS (${\mathbf{\Phi}}_r$ and ${\mathbf{\Phi}}_t$), which the complexity involved is $\mathcal{O}\left({{\left(L+1 \right)}^{3.5}}+{{L}^{3.5}}\right)$.  Therefore, the overall computational complexity of the {AIOA} is calculated as $\mathcal{O}\left( I\left( KM+MN+{{\left( L+1 \right)}^{3.5}}+{{L}^{3.5}} \right) \right)$, where $I$ denotes the total number of iterations.

\section{Numerical Results and Discussion}\label{sention_6}
\subsection{System Deployment and Parameters Setting} 
\begin{figure}
	\centering
	\includegraphics[width=1.0\columnwidth]{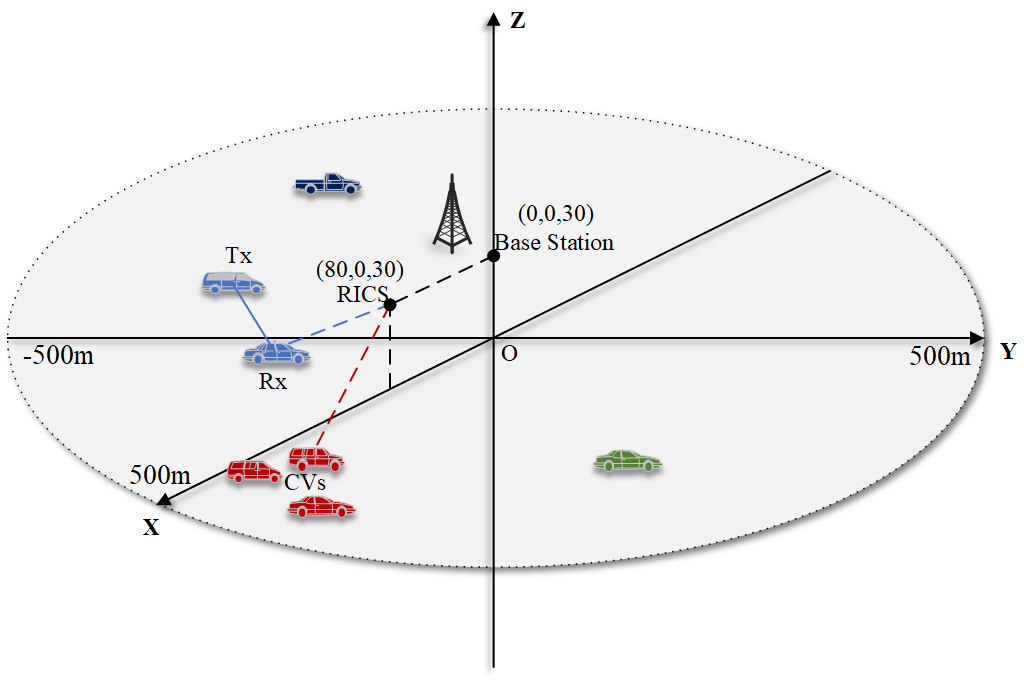}
	\caption{The simulated RICS-based autonomous driving scenario on a 500m circular track. Where, red vehicles represent CVs and blue vehicles represent V2V pair.}
    \label{location}
\end{figure}

We evaluate the performance of the proposed {AIOA} algorithm and the amplitude adjustment function of the RICS. The simulation scenario is set in a three-dimensional spatial coordinate system, where the ${x-y}$ plane is a circular field with a radius of $500$ meters. As shown in Fig.~\ref{location}, the BS is located at the origin $(0,0,30)$, while the RICS is positioned 80 meters away from the BS, i.e., $(80,0,30)$. Meanwhile, the vehicles are distributed on the road according to a Poisson process. $M$ vehicles are randomly selected as CVs to form a V2I link with the base station. V2V pairs are then constituted by selecting each vehicle with its nearest neighboring vehicle. The specific simulation parameters are shown in Table~\ref{parameters}.

\begin{table}[hbt]
	\centering
	\caption{Simulation Parameters}
	\label{parameters}
	\small
	\begin{tabular}{l | l || l | l}
		\noalign{\global\arrayrulewidth=0.3mm}
		\hline
		\textbf{Parameter}  &\textbf{Value } &\textbf{Parameter}  & \textbf{Value } \\
		%\noalign{\global\arrayrulewidth=0.1mm}
		\hline
		 $L$   & 30 &  $M$   &10\\
		 $N$  &10 &  $s_m$     &15 Mbits \\
		$f_m$  & [1, 5] GHz & $F$ & 10 GHz \\
		 $P_m$   &28dBm  & $P_t$    &     23 dBm\\
		${\epsilon}$ &  $10^{-6}$ & $P_{outage}$        &0.01\\
		  $\delta $    &    $10^{-3}$ & $\kappa $   &    4\\    
		${\gamma}_{th}$ &2 bps/Hz& $\alpha$ &2.5\\       
		$C_0$ &-30dB &   $W{{\xi }_{0}}$ &    -110 dBm\\
            $\lambda$ &0.8&   $A_B(Q)$ &0.9 \\
		\noalign{\global\arrayrulewidth=0.3mm}
		\hline
	\end{tabular}
\end{table}	

\subsection{Convergence Performance} 
Fig.~\ref{sim1} shows the convergence of the proposed AIOA algorithm with 30 and 64 RICS elements (denoted as $L$), respectively. It clearly shows that the proposed AIOA in both two cases converged when the number of iterations is 5 and 6. Moreover, we observe that the proposed AIOA algorithm with $L=64$ outperforms the case with $L=30$, which indicates that the average safety coefficient of each CV can reach around ${76\%}$ with 30 RICS elements and $85.5\%$ with 64 RICS elements.

However, the average safety coefficient of $L=64$ is only $8\%$ higher than $L=30$, which means as the number of elements increases, the improvement of the safety factor is not significant, and the marginal benefit of increasing the elements is lower. This may be due to the proposed global optimization algorithm AIOA improving the overall safety coefficient by dynamically adjusting the offloading ratio. This also means that our algorithm can flexibly choose the offloading ratio according to different situations. This optimization strategy pays more attention to the improvement of the overall performance.  Even in the case of a fragile V2I link, the overall safety coefficient can still be kept at a relatively high range.

%\begin{figure}[h]
%	\centering
%	\includegraphics[width=1.0\columnwidth]{converge.pdf}
%	\caption{Convergence of the proposed Algorithm(\textbf{AIOA}).}
%    \label{sim1}
%\end{figure}

\subsection{Impact of Different Optimization Methods} 
In this part, we proposed three schemes as benchmarks for alignment with our proposed {AIOA}: \vspace{-0.8mm}
\begin{itemize}
\item \textbf{Random offloading}: the offloading ratio $\bm{\rho}$ varies within $[0, 1]$.
\item  \textbf{RICS with the random $\bm{{\Phi}_{x}}$}: the RICS refraction and reflection phase shifts vary in [0, $2{\pi}$].
\item $\textbf{Random spectrum sharing}$: V2V links randomly sharing the spectrum of V2I link occupancy. Then, we explore the change in transmission power of CVs $P_{m}$ on the safety coefficient. We selected the RICS elements $L=30$ and controlled the $P_{m}$ varying from 20dBm to 40dBm.
\end{itemize}

\begin{figure*}[t] 
\centering
\subfloat[]
{
\includegraphics[width=0.65\columnwidth]{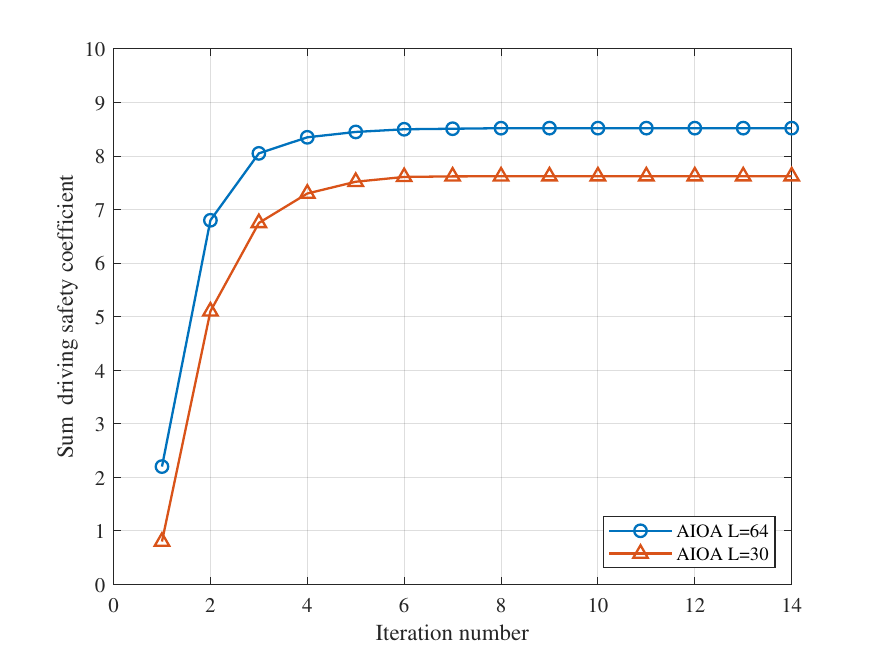}
\label{sim1}
}
\hfill
\subfloat[]
{
\includegraphics[width=0.65\columnwidth]{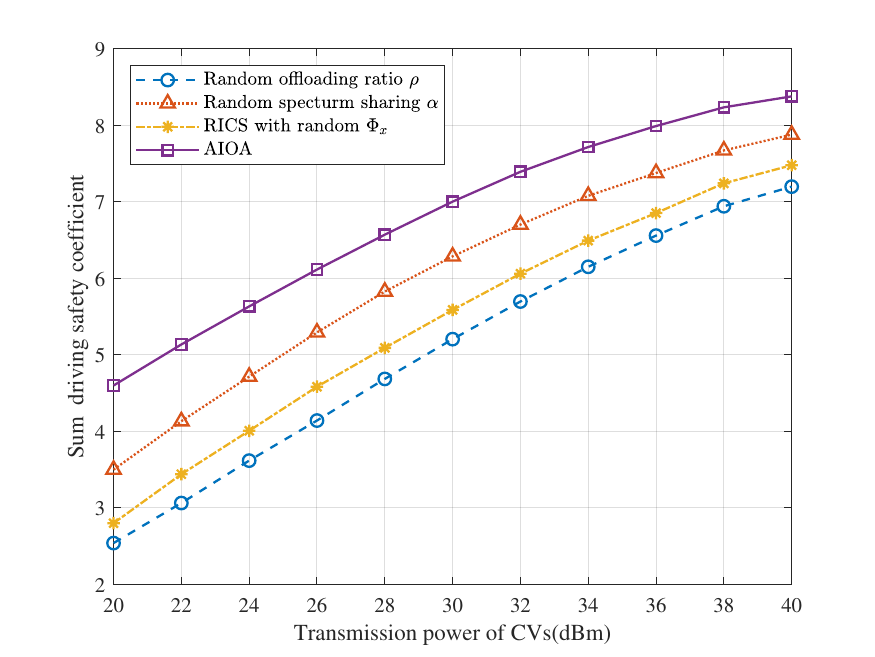}
\label{transmmissionpower}}
\hfill
\subfloat[]
{
\includegraphics[width=0.65\columnwidth]{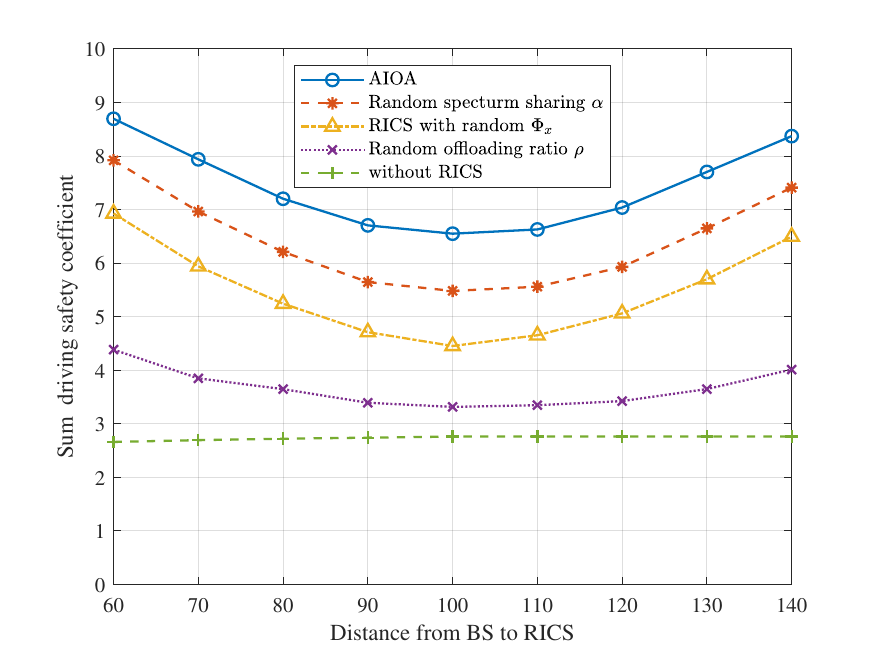}
 \label{dist}}
\caption{Convergence of the proposed AIOA algorithm is shown in (a). The sum driving safety coefficient with a varying transmission power of CVs is shown in (b), where $N=5$. The sum driving safety coefficient with varying distances from BS to RICS is shown in (c).}
\end{figure*}

%\begin{figure}[h]
%	\centering
%	\includegraphics[width=1.0\columnwidth]{power_Sm.pdf}
%	\caption{Sum driving safety coefficient with varying transmission power of CVs.}
%	\label{transmmissionpower}
%\end{figure}

Fig.~\ref{transmmissionpower} shows the proposed AIOA scheme compared with the other three benchmark changes with $P_{m}$. Here, we selected the RICS elements $L=30$, $P_{t}$ is set as 22dBm, and controlled the $P_{m}$ varying from 20dBm to 40dBm. The simulation results show that as the transmission power of CVs $P_{m}$ increases, the sum of the driving safety coefficient increases. This is because the effect of transmission power on the $R_{B}^{m}$ also indirectly improves the safety of the system. Besides, we can clearly see that the proposed AIOA can offer nearly [34\%, 9\%, 25\%] performance gain compared to the other three schemes. We can also see that with the increase of $P_{m}$, the growth rate of the sum safety coefficient begins rapidly and then decelerates, and gradually tends to saturation. This may be because as the transmission power increases, the interference in the system also increases, which has a negative impact on the increase of the safety coefficient.

%The simulation results show that as the $P_{m}$ increases, the safety coefficient increases sharply. This is because the effect of transmission power on the ${R_{B}^{m}}$ also indirectly improves the safety of the system. Besides, we can also see that with the increase of $P_{m}$, the growth rate of the sum safety coefficient is begin rapidly and then decelerate, and gradually tends to saturation. In this range, our \textbf{AIOA} is always due to the other three comparison algorithms, reflecting the superiority of our algorithm.
%\begin{figure}[h]
%	\centering
%	\includegraphics[width=1.0\columnwidth]{distance_plot.pdf}
%	\caption{Sum driving safety coefficient with varying distance from BS to RICS.}
 %   \label{dist}
%\end{figure}

\begin{figure*}[t] 
\centering
\subfloat[]
{
\includegraphics[width=0.65\columnwidth]{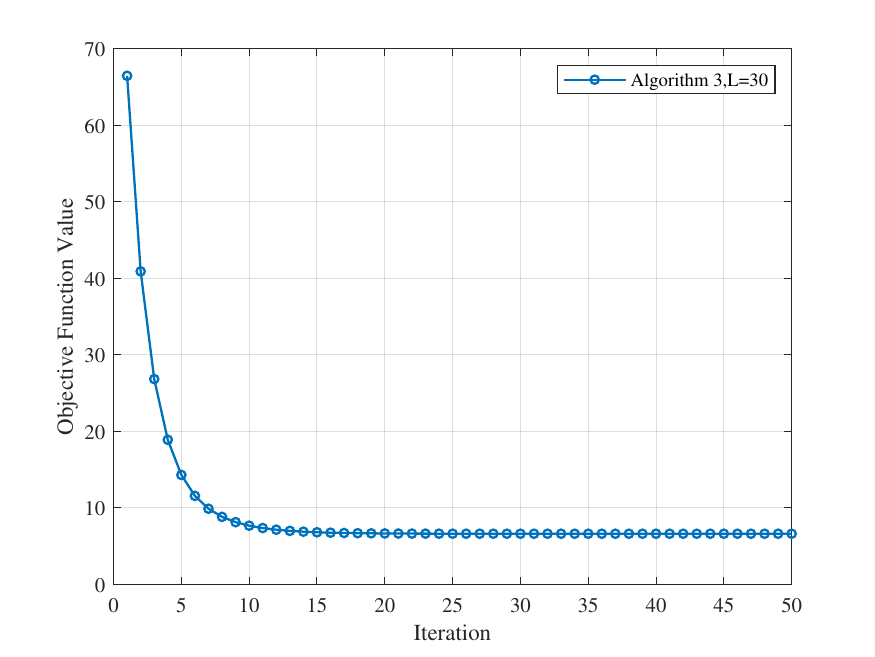}
 \label{Psi}
}
\hfill
\subfloat[]
{
\includegraphics[width=0.65\columnwidth]{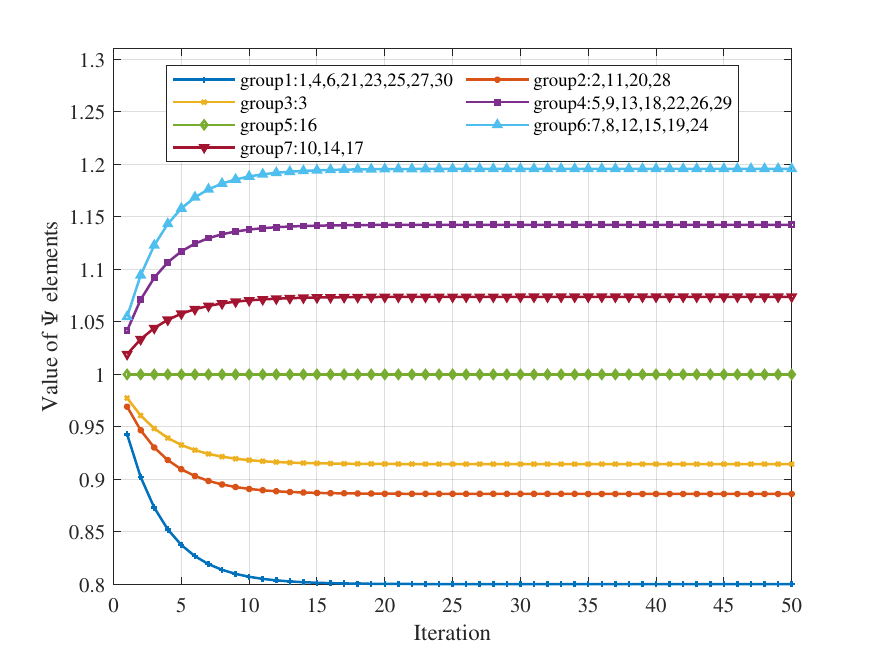}
\label{Obj}}
\hfill
\subfloat[]
{
\includegraphics[width=0.65\columnwidth]{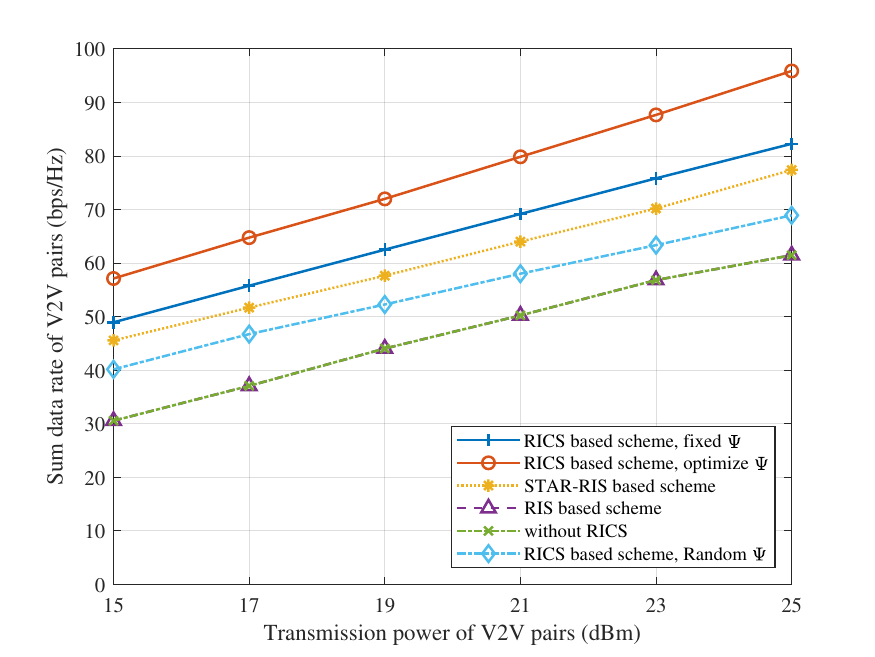}
\label{Rn_plot}}
\caption{ Convergence behavior of the objective function is shown in (a). The convergence behavior of amplitude adjustment factor $\bm{\Psi}$ is shown in (b). The available V2V data rate with varying $P_{t}$ for $N=10$ is shown in (c).}
\end{figure*}

\begin{figure*}[t] 
\centering
\subfloat[]
{
\includegraphics[width=0.9\columnwidth]{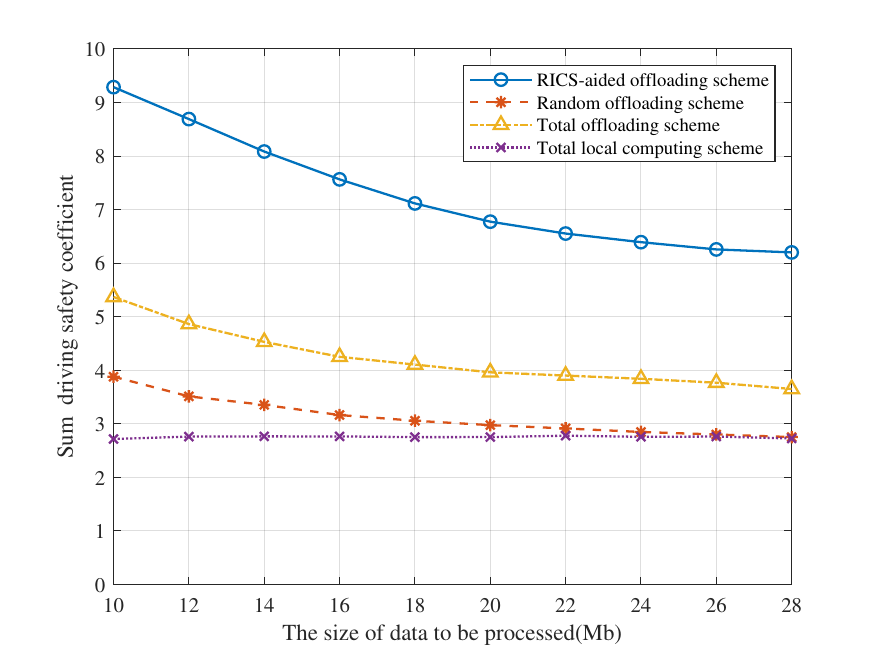}
 \label{plot_sm}
}
\hfill
\subfloat[]
{
\includegraphics[width=0.9\columnwidth]{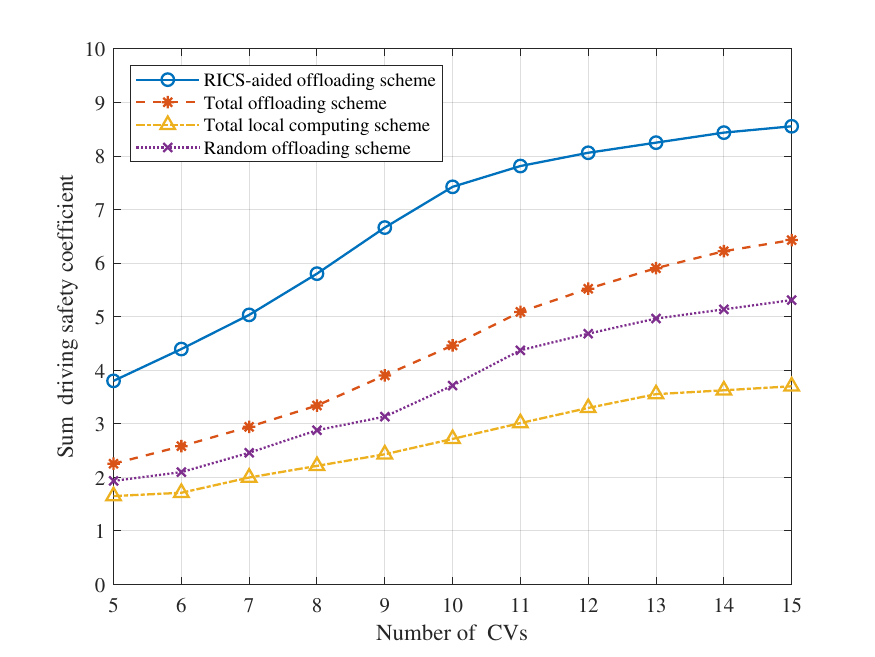}
\label{off_scheme}}
\caption{Sum safety coefficient of CVs with varying value of input data size $s_{m}$ is shown in (a). The sum safety coefficient with varying number of CVs is shown in (b).}
\end{figure*}

The location of the vehicles and the infrastructure in the system may affect the performance of the sum safety coefficient. Thus, we explore the effect of the distance between RICS and BS varying from 60 to 140 meters on the safety coefficient. Here, the RICS and BS are located at $(d_{RICS},0,30)$, $(0,0,30)$, where, the distance between BS and RICS is denoted as $d_{RICS}(m)$. Besides, we define the location of the central CV as $(200,0,0)$. According to Fig.~\ref{dist}, the safety coefficient decreases and increases after increasing $d_{RICS}$, and is lowest at the distance of 100m. This trend may be explained as follows. When the RICS is located at the midpoint of the BS and central CV, the PL(BS-RICS-CV) becomes maximum because ${x}^{BS}_{RICS}=x^{RICS}_{cCV}=\frac{1}{2}x^{BS}_{cCV}$ according to average inequality, which leads to the weak signal from BS to central CV. However, when the distance is too far, the power of the reflected signal from RICS reaching the BS decreases, leading to a reduction in the sum safety coefficient of the CVs. So, the rational deployment of RICS is intuitively important for enhancing system security. Furthermore, the plots also show the necessity of joint optimizing the three variables because our AIOA compared these three schemes to improve the sum safety coefficient of [104\%, 19\%, 45\%] at $d_{RICS}=100$m.

\subsection{The Impact of the Amplitude Adjustment Factor}
Fig.~\ref{Psi} illustrates the trend of the objective function value with the number of iterations. In the first 10 iterations, the value of the objective function decreases rapidly, indicating that the amplitude adjustment factor $\bm{\Psi}$ is rapidly approaching its optimal value. As the number of iterations increases, the decrease of the objective function value gradually slows down and eventually stabilizes, indicating that the algorithm has converged to the local optimal solution. Specifically, the objective function of our algorithm converges to 6.6199 at the $44$\textit{th} iteration. It is worth noting that the objective function value converges to a fixed minimum value instead of 0. This may be due to the nonlinear relationship between the objective function and the amplitude tuning factor, and the coupling between different elements of the $\bm{\Psi}$ vector. Therefore, it is difficult to achieve an objective function value of 0 by adjusting $\bm{\Psi}$ alone. However, a suitable amplitude adjustment factor can effectively mitigate some of the interference and thus enhance the data rate of the V2V pair, as will be shown in Fig.~\ref{Rn_plot}.
 
 Fig.~\ref{Obj} illustrates the changes in amplitude adjustment factors value of $\bm{\Psi}$ with the number of iterations. In order to analyze the convergence behavior of $\bm{\Psi}$ values in the gradient descent method more clearly, we merge the $\bm{\Psi}$ with similar final results. Specifically, we set a threshold value of 0.05 to categorize those with final $\bm{\Psi}$ differences less than this threshold into the same group, and finally, we categorize them into seven groups. It can be observed that the value of $\bm{\Psi}$ gradually stabilizes with the number of iterations and the performance converges nearly after 20 iterations. Besides, the value of L adjustment factor $\bm{\Psi}$ range for $[0.8,1.2)$. Different RICS adjustment factors for the corresponding elements may amplify or weaken the amplitude, which may be due to the different distances between the CVs and the V2Vs, resulting in different levels of signals that need to be used to mitigate the interference. The signal transmitted through the RICS serves more of what is left of the CVs, and therefore needs to be optimized by the $\bm{\Psi}$ to target the interference. 
 
 To further highlight the role of the amplitude adjustment factor, we set the RICS-based scheme for the optimized $\bm{\Psi}$ configuration to compare with five benchmark schemes:
\begin{itemize}
\item $\textbf{RICS with fixed $\bm{\Psi}$}$: the amplitude adjustment factors $\bm{\Psi}=1.2$.
\item $ \textbf{RICS with random $\bm{\Psi}$}$: the amplitude adjustment factors $\bm{\Psi}$ varies in [1,2].
\item $\textbf{STAR-RIS}$: without amplitude adjustment.
\item $\textbf{RIS}$: In the network, we only consider the tranditional RIS-assisted links and direct links.
\item $\textbf{Without RIS}$: the considered network does not contain assistance from RICS, which means we only jointly optimize the offloading ratio and spectrum sharing.
\end{itemize}

Fig.~\ref{Rn_plot} demonstrate the performance of the six schemes and the transmission power of V2V pairs varying from $15$dBm to $25$dBm, where the $P_{m}$ is set as $23$dBm. As $P_{t}$ increase, all six schemes rise, which means that increasing $P_{t}$ helps to boost the sum data rate of V2V pairs. Furthermore, the performance of the RICS-based scheme, optimized $\bm{\Psi}$ surpass ${}$ better compared to the other five schemes. At the same time, the based scheme fixed $\bm{\Psi}$ outperforms random $\bm{\Psi}$. This result indicates that the amplitude adjustment factors can be effectively configured to mitigate the interference and improve the sum data rate of V2V. Besides, the STAR-RIS-based scheme performs better than the random $\bm{\Psi}$, which reflects the importance of reasonable configuration, the random amplitude adjustment factor may be counterproductive, increasing the interference of $Rx$. The performance of the RIS-based scheme and without RICS is the same because neither of the two schemes has an RICS-assisted refraction link.

\subsection{Impact of Different Offloading Schemes}
We proposed three schemes for comparison with our RICS-aided offloading scheme.  %: \textbf{STAR-RIS aided Rn scheme}, \textbf{RICS-aided Rn $\bm{\Psi} = 1.2$ scheme}, \textbf{RICS-aided Rn $\bm{\Psi} = 1.5$ scheme}, \textbf{without RIS aided scheme}
In order to validate the RICS-aided offloading strategy, we chose the optimized $\bm{\Psi}$ to explore the performance of different offloading schemes:
\begin{itemize}
\item $\textbf{Total offloading}$: The value of ${\rho}$ all taken as 1.
\item $ \textbf{Total local computing}$: The value of ${\rho}$ all taken as 0.
\item $\textbf{Random offloading}$: ${\rho}$ is taken as 0 or 1.
\end{itemize}

\vspace{-0.5mm} %The simulation results show that as the $P_{t}$ increases, the sum rate of V2V pairs increases. Besides, we can see that the performance of RICS-aided Rn, $\bm{\Psi} = 1.2$ is optimal. And next is STAR-RIS aided Rn. But as the $\bm{\Psi}$ increases, RICS-aided Rn scheme doesn't perform well. This may be due to the fact that a single amplification does not cancel better with the interference, and proper amplification is what allows the interference of the V2V pairs to be effectively cancelled thus increasing the rate. From the figure 4, $\bm{\Psi} = 1.2$ is better for assisting in boosting the sum rate of V2V pairs, which also demonstrates the advantages of our proposed structure RICS.
Fig.~\ref{plot_sm} illustrates the performance of the sum safety coefficient of the three schemes varying from the input processing data $s_{m}$. Except for the total local computing scheme, the other three schemes decrease as the input data $s_{m}$ increases and gradually plateaus. In addition, the performance of the RICS scheme exceeds the other three schemes. Both total local computing scheme and total offloading scheme are subpar, with high local computation latency and inability to use BS resources in the case of the former, and high transmission latency and potential bottleneck of BS computation resources in the case of latter. Therefore, our RICS-aided offloading scheme is able to balance the two aspects well, thereby improving the safety coefficient. The reason the total local computing scheme maintains a relatively stable value is that the input data size $s_{m}$ only affects offloading latency, and this strategy incurs only local latency, therefore the safety coefficient will not show significant changes.

Fig.~\ref{off_scheme} explores the influence of the number of CVs on the sum safety coefficient under different offloading strategies. 
Simulation results show that for our prposed AIOA and total offloading at the BS scheme, the sum safety coefficient first rises and then flattens out. Especially for AIOA, we can discover that the average safety coefficient of CVs achieves the maximum value when it is 10, that is to say, the $M/N=1$. This is possibly due to the fact that when the number of CVs is in the range of 5-10, the system has enough capability to bear more CVs to perform offloading while mitigating interferences suffered at the V2V pairs. This will intuitively respond by increasing the ${R}_{B}^{m}$, which further enhances the security of the system. The random offloading scheme, on the other hand, is reflected in a slight oscillation of the safety coefficient with the growth rate of the CVs, due to the uncertainty of whether it is offloaded locally or at the BS. But in terms of overall trends, our AIOA is always better than the other two modes. And the total offloading scheme is slightly better than the total local computing scheme. This may be because the powerful computational capabilities of the BS enable faster processing of tasks and can compensate for the additional transmission delay. Furthermore, it can reduce the computational load on the vehicles themselves to lower energy consumption, thus improving the safety coefficient.

\section{Conclusion}\label{sention_7}
In this paper, we presented a novel RICS-aided computation offloading framework for enhancing the safety coefficient in autonomous driving networks. The proposed design problem, which was non-convex and solved via an alternating optimization algorithm, involved optimizing the offloading ratio, spectrum sharing, as well as the RICS refraction and reflection coefficients, while satisfying the V2V outage probability. To explore the optimal signal adjustment factor for configuring the RICS that works in  AC mode, we utilized the gradient descent algorithm, which was shown to achieve excellent convergence. Our extensive numerical investigations showcased 
% is motivated by an autonomous vehicle accident and a novel offloading framework is proposed to minimize the inference error subject to latency constraint. The optimal offloading probability and the pre-braking probability are analyzed to 
that the proposed RICS-aided offloading framework, not only achieves high inference accuracy of CVs, but also perfectly mitigates interference at V2V pairs.

%{\appendix[Proof of the Zonklar Equations]
%Use $\backslash${\tt{appendix}} if you have a single appendix:
%Do not use $\backslash${\tt{section}} anymore after $\backslash${\tt{appendix}}, only $\backslash${\tt{section*}}.
%If you have multiple appendixes use $\backslash${\tt{appendices}} then use $\backslash${\tt{section}} to start each appendix.
%You must declare a $\backslash${\tt{section}} before using any $\backslash${\tt{subsection}} or using $\backslash${\tt{label}} ($\backslash${\tt{appendices}} by itself
% starts a section numbered zero.)}


\end{document}